\documentclass[reprint,amsmath,amssymb,aps]{revtex4-2}
\usepackage{graphicx}
\usepackage{dcolumn}
\usepackage{bm}
\usepackage{float}
\usepackage{amsmath}

\begin{document}

\title{Automating Blueprints for Colloidal Quasicrystal Assembly}

\author{Diogo E. P. Pinto$^{1}$, Petr \v{S}ulc$^{2,3,4}$, Francesco Sciortino$^{1}$, and John Russo$^{1}$}

\affiliation{$^{1}$ Dipartimento di Fisica, Sapienza Universit\`{a} di Roma, P.le Aldo Moro 5, 00185 Rome, Italy \\ $^{2}$ School of Molecular Sciences and Center for Molecular Design and Biomimetics, The Biodesign Institute, Arizona State University, 1001 South McAllister Avenue, Tempe, Arizona 85281, USA\\$^{3}$ Center for Biological Physics, Arizona State University, Tempe, Arizona, USA \\ $^{4}$ TU Munich, School of Natural Sciences, Department of Bioscience, Garching, Germany}

\begin{abstract}

One of the frontiers of nanotechnology is advancing beyond the periodic self-assembly of materials. Icosahedral quasicrystals, aperiodic in all directions, represent one of the most challenging targets that have yet to be experimentally realized at the colloidal scale. Previous attempts have required meticulous human-designed building blocks and often resulted in interactions beyond current experimental capabilities. In this work, we introduce a framework for generating experimentally accessible designs that self-assemble into quasicrystalline arrangements. We present a design for icosahedral DNA-origami building blocks and demonstrate, through molecular simulations, their successful assembly into the target quasicrystalline structure. Our results highlight the feasibility of using automated design protocols to achieve complex quasicrystalline patterns, paving the way for new applications in material science and nanotechnology.

\end{abstract}

\maketitle

\section{Introduction}

One of the greatest achievements of nanotechnology is the self-assembly of crystalline structures with tailored properties, such as photonic crystals with omnidirectional band gaps \cite{Posnjak2024, Liu2023, Macfarlane2011, Jones2015}. The task of creating building blocks for crystal assembly is considerably facilitated by translational symmetry, which allows for targeting only a small number of particles (the unit cell) to direct the assembly pathway \cite{Romano2011, Wang2012}. The biggest challenge encountered in crystal assembly is the need to select one and only one of the possible polymorphs (competing structures) as well as to avoid kinetic traps that arise from the bonding degeneracy of the building blocks. Several approaches have been proposed to tackle these difficulties, from geometric design \cite{Videbaek2024, Flavell2023} to optimization methods \cite{King2024, Coli2022}.

The problems faced in self-assembly become considerably more challenging when the target structures are aperiodic (as in quasicrystals) or when the unit cell becomes arbitrarily large (as in quasicrystal approximants, which have short-range quasicrystalline orientational order but long-range periodicity) \cite{Goldman1993}. Theoretical proposals have focused on directional interactions that enforce quasicrystalline orientational symmetry. These interactions have either an entropic origin \cite{Engel2015}, when controlled by particle shape, or an energetic origin, when controlled by selective bonds \cite{Noya2021}. Recently, experimental realizations of DNA-coated nanoparticles have successfully assembled the dodecagonal quasicrystal, which has 12-fold symmetry along one of its planes and is periodic along the perpendicular axis \cite{Zhou2024}. Other approaches have also achieved successful assemblies at the microscale, particularly in 2D \cite{Plati2024}.

One of the unmet goals is the laboratory realization of a fully aperiodic colloidal structure in 3D, such as the icosahedral quasicrystal. An important theoretical milestone in this direction was achieved by Noya and collaborators \cite{Noya2021}, who designed a patchy particle model capable of assembling the icosahedral quasicrystal. The work utilized torsional interactions on top of coloured patch specificity to nucleate the correct structure.
Lee and Glotzer \cite{Lee2022} were able to obtain five-fold and icosahedral twinned clusters with purely repulsive interactions, utilizing the concept of entropic bonds as a way to drive hard core particles to assemble into specific structures, including quasicrystals \cite{Engel2015}. In this case, the assembly is completely driven by entropy but requires particles to have specific shapes to properly form the patterns seen in the quasicrystals, and the high density requirement for the assembly to be driven by entropy is still prohibitive for many applications. Recently, it has been shown that a system of hard spheres with two characteristic sizes in 2D is able to nucleate a quasicristalline structure due to the high configurational entropy of the quasicrystal \cite{Fayen2024}. Moreover, these structures are able to accommodate a significant number of defects and still show the intended symmetries. This is observed due to the preference of the system to assemble the geometrical patterns that characterize the quasicrystal (square and triangles).

Here, we propose a new solution derived from the SAT-assembly framework for the self-assembly of the icosahedral quasicrystal and its approximants. SAT is an automated strategy that converts the assembly problem into a Boolean Satisfiability Problem, which also allows for the suppression of competing structure formation~\cite{Russo2022}. Our design utilizes currently available building blocks, recently used to assemble in the lab the pyrochlore lattice \cite{Liu2023}. These building blocks consist of icosahedral DNA origami \cite{sigl2021programmable, Mosayebi2017, Lee2022a, Jun2021, Rothemund2006} with single-stranded overhangs. The nucleotide sequence of the overhangs encodes the specificity of the interaction, with bonds forming only when complementary strands on different icosahedral origami come together. This design relies solely on bond specificity and does not require additional interactions (such as torsional interactions), which are more challenging to design in DNA origami, especially if it has many interacting sites, each of which needs to be torsionally restricted to allow only specific torsional angle between interacting particles, such as in the solution presented in Ref.~\cite{Noya2021}.
As we alluded above, the SAT-assembly method also allows us to incorporate the avoidance of common competing structures, i.e., crystal structures that would otherwise form with a subset of the components, into the solution. After deriving the solution, we map it to a patchy particle model \cite{Sacanna2011, Wang2012, Wang2015, Rovigatti2022, ravaine2017synthesis, pawar2010fabrication} and run molecular simulations to verify that it correctly assembles the target structure. Finally, we demonstrate how the solution can be implemented using icosahedral DNA origami, which is based on previously experimentally realized designs \cite{Liu2023, Zhang2022}.

\section{Results and discussion}

\subsection{SAT-assembly for quasicrystals}

We approximate the icosahedral DNA origami by a patchy particle where each patch corresponds to a DNA overhang. If we consider that each DNA sequence corresponds to a patch colour, then the inverse design problem becomes a colouring problem. This can be approached by finding the number of colours, the way these are distributed through the patches, and the colour interaction matrix such that it only assembles the target quasicrystal. For simple structures such process can be done by hand, but as the complexity of the structure increases, so does the difficulty of the problem. Here, we use the previously developed SAT-assembly algorithm \cite{Russo2022} to solve this task.

The SAT-assembly framework has so far been applied to crystals and finite-size structure with the same local environment, i.e. with the same geometry of the building block~\cite{Russo2022, Bohlin2023, Pinto2023, Romano2020}. Here, we extend the framework to handle many local environments, such as those found in quasicrystals and their approximants. This allows us to automatically generate blueprints for their assembly. SAT-assembly translates the self-assembly problem into a colouring problem, where the directional interaction of the building blocks is encoded in patches (attractive spots on the surface) and the specificity in their colouring. Each building block is distinguished not only by the geometrical arrangements of the patches but also by their colouring. In the following, the number of species - which corresponds to the number of unique building blocks - is indicated with $N_p$, and the total number of colours with $N_c$. 

The resulting SAT algorithm goes through the following steps:\\
\emph{i}) Define  the inter-particle bonds in the target structure and assign to each particle a unique local environment (based on the topology of the bonding pattern). Nearest neighbour particles, defined as particles with a relative distance smaller than a cutoff threshold that includes the first three peaks of the radial distribution function, are considered bonded.
  \\
\emph{ii}) Generate Boolean clauses that enforce the topology of the target structure and the exclusion of competing structures. The SAT clauses (\textit{ii}) that enforce the topology are described in detail in the \textit{Supporting Information}.\\
\emph{iii}) Choose how many species $N_p$ (i.e. different building blocks) to use for each environment, and the total number of colours $N_c$. The minimum value of $N_p$ and $N_c$ are such that the SAT-problem is satisfiable; increasing their number increases the constraints on the self-assembly pathway but slows down its kinetics and is more challenging to realize experimentally. The right choice is usually a balance of these factors.\\
\emph{iv}) Use SAT-solvers \cite{Een2005} to find a solution which consists in a colouring interaction table, the patch colouring of each species, and their position in the target structure. SAT-solvers are able to efficiently solve millions of clauses with millions of different boolean variables which is the typical range of the problems addressed here.\\
\emph{v}) Run molecular simulations with the discovered design to assess its ability to reach the target structure.  We run Monte Carlo simulations (\textit{v}) with the Kern-Frenkel potential~\cite{bol1982monte,Kern2003}, which is a minimal model still able to reproduce the self-assembly behavior of DNA-origami with single stranded overhangs \cite{Liu2023}. Simulations are essential to check whether the proposed design assembles into the desired structure.\\
\emph{vi}) Simulations that do not form the target structure can be divided in two categories: in the first group, the self-assembly has been hindered by a competing structure; in the second, the kinetics of aggregation of the designs has favoured amorphous aggregates. Within SAT-assembly the first problem can be avoided by including additional Boolean clauses which forbid the formation of the competing topology (going back to step \emph{ii}).The second problem can be overcome by finding a new solution with an increased number of colours (going back to step \emph{iii}). The ability to explicitly exclude the formation of competing structures is one of the main advantages of the SAT-assembly framework.

\subsection{Icosahedral quasicrystal approximant}

\begin{figure*}[t]
	\includegraphics{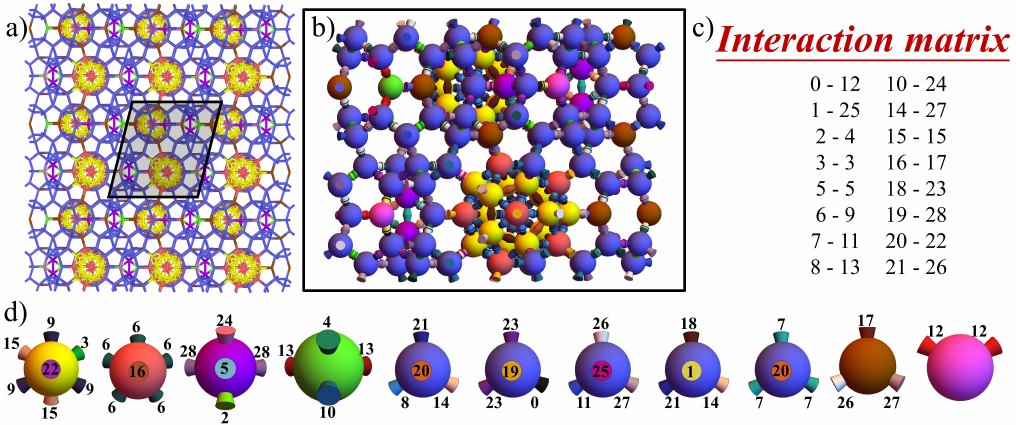}
	\caption{\label{fig1} Schematic representation of the target structure and building blocks for the self-assembly of an icosahedral quasicrystal approximant. In a) is the target structure used, the $3/2$ icosahedral quasicrystal approximant, with only the bonds between the particles drawn. Each colour represents an environment, where patch geometry is different. In b) is a schematic representation of the unit cell with $284$ particles and the respective coloured patches corresponding to the $11$ species and $29$ colours design. c) and d) show the details of the design used, particularly the different patch species and the corresponding interaction matrix between the patch colours. In this case, six of the species correspond to unique environments while the remainder five species correspond to the same 4-patch environment (blue).}
\end{figure*}

Quasicrystal approximants are periodic structures that have similar properties to those shown by the ideal quasicrystal. For example, an approximant shows a similar diffraction pattern to the one of the ideal structure~\cite{Goldman1993}. The ideal quasicrystal can be generated by using the cut-and-project method on a 6D hypercube. In this method, one can separate the higher-dimensional space into two orthogonal subspaces, the perpendicular and the physical spaces. The cut is done by projecting a 6D body-centred hypercubic lattice onto the perpendicular space and removing the points outside a dodecahedron occupation domain. The occupation domain choice fixes the point group symmetry of the quasicrystal. The subset of points inside the occupation domain can then be projected again from 6D onto the physical plane to obtain the particle positions for the ideal quasicrystal. The values used in the first projection to perform the cut fixes the structure that is projected onto the physical space. If an irrational cut is used, e.g. using the golden ratio, then the ideal quasicrystal is generated, if instead two successive Fibonacci sequence numbers $q/p$ are used, the periodic $q/p$ approximant is generated. Fig.~\ref{fig1}a) shows a section of the $3/2$ icosahedral quasicrystal approximant, generated from the cut-and-project method, oriented along the pseudo-five-fold symmetry axis. Particles are considered to be bonded if their relative distance is smaller than $1.7$ in the representation obtained from the cut-and-project method. This cut off values separates first and second neighbours. We use this to define the patch geometries which can be organized into $7$ unique environments (Fig.~\ref{fig1}d).

In the simulations, we consider a system composed of $N$ patchy particles in a cubic box of length $L$. Each particle is characterized by a hard core of radius $\sigma$ and $N_{patches}$ patches located on its surface, and it interacts with other particles according to the Kern-Frenkel potential~\cite{bol1982monte,Kern2003}. In Fig.~\ref{fig1}b) we show a schematic representation of the unit cell of the $3/2$ icosahedral quasicrystal approximant where each particle is given a colour corresponding to their environment, and the different environment patch geometries are shown. The results shown below were obtained via Monte Carlo simulations using aggregation-biased moves~\cite{Rovigatti2018}. More details regarding the simulation details can be found in the \textit{Supporting Information}.

We consider that each patch can have a given colour $x_c$ between $0\leq x_c< N_c$, where $N_c$ is the total number of distinct colours. These colours can be distributed onto the patches in specific arrangements, each unique sequence can be considered a particle specie $x_p$, thus $1\leq x_p\leq N_p$, where $N_p$ is the total number of distinct species. SAT is then used to find if a given combination of $N_p$ and $N_c$ can satisfy the target structure, e.g. if it satisfies all the topological constraints, and a solution/design is calculated which can be used to prepare the composition of the system. In the \emph{Supporting Information} we go into more detail on the different constraints (clauses) used for SAT. The particles shown in Fig~\ref{fig1}c) and d) result from a design with $11$ species and $29$ colours. The $11$ species corresponds to at least one of each environment plus other possible colour variations for each. In this design, the blue 4-patch environment represents five of the species with different patch colour arrangements, while the remainder six species correspond to unique environments with different patch geometries. 

Through the SAT formulation we can recursively generate an array of different coloured designs with different colour arrangements. Moreover, we can constrain these designs to exclude specific competing structures. Previous works have noted that similar patch geometries can nucleate the $BC8$ crystal which directly competes with the icosahedral quasicrystal~\cite{Noya2021}, thus requiring torsional interactions to properly assemble the target structure. In our framework, we eliminate this minima by constraining the patchy particle design to not satisfy the topology of the $BC8$ crystal unit cell (see \textit{Supporting Information}). The SAT algorithm guarantees that the patchy particle design shown in Fig.~\ref{fig1}c) and d) with $11$ species and $29$ colours does not satisfy the $BC8$ crystal topology.

\begin{figure}[t]
	\includegraphics{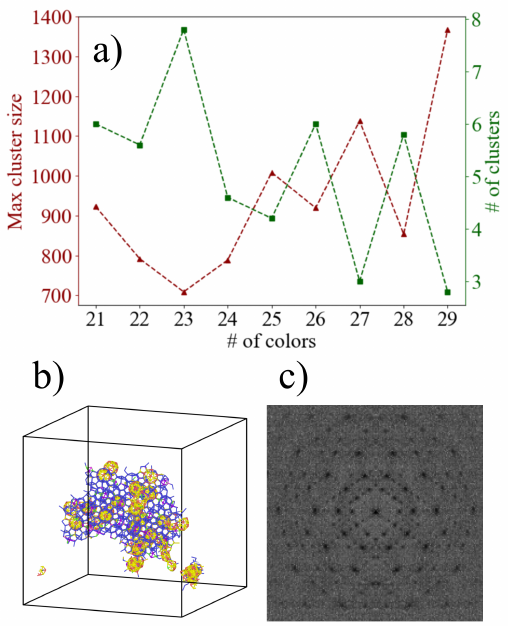}
	\caption{\label{fig2} Seeding and direct nucleation of the icosahedral quasicrystal approximant. In a) are shown results for the icosahedral quasicrystal approximant, where the structure is seeded using a single unit cell of $284$ particles at the center of the simulation box. The system is simulated for $\rho=0.05$ and $T=0.09$, and $\Delta t=2\cdot 10^7$. We simulate this system using nine different coloured designs and show how the maximum cluster size and the total number of clusters changes as a function of the total number of colours. The simulation box contains $N=2680$ (including the initial seed) particles and we averaged over $5$ different samples. In b) is shown a cluster formed from a direct nucleation trajectory where only the bonds between particles are drawn. The direct nucleation simulation was done for $T=0.096$, $\rho=0.1$, $N=5680$, and $\Delta t=2\cdot 10^8$. The cluster presented has size $N=2221$ and is oriented along the pseudo-five-fold symmetry axis. The corresponding diffraction pattern along this axis is shown in c).}
\end{figure}

For the following results we have fixed the total number of species to $11$, which corresponds to the distribution of environments shown in Fig.~\ref{fig1}d). Moreover, we also fix the specie of each particle in the topology of the structure. Due to the high degree of complexity in the topology of the quasicrystal approximant, fixing these variables aids the SAT-solver to reach a solution in reasonable time. To reach this value of species we have considered the bonded neighbours of each particle and kept only the unique arrangements of bonds between environments which cannot be obtained from simple rigid-body rotations of the particles. Then, through SAT, we can use different values for the total number of colours and create a pool of designs (patch colour arrangements and colours interaction matrix) that can be tested for assembly. Fig.~\ref{fig2} shows the results for nine different designs, going from $21$ to $29$ colours, where we have seeded one unit cell of the $3/2$ icosahedral quasicrystal approximant (which contains $284$ particles) in a cubic box with a total of $N=2840$ particles. We run simulations at $T=0.09$ (with the energy of the attractive interaction set to $\epsilon/k_B=1$) and $\rho=0.05$ (with the diameter of the individual particles set to $\sigma=1$) to minimize the nucleation of other structures while allowing the seed to grow. We keep the same ratio of species as the one of the unit cell. We find that the size of the largest cluster increases with the total number of colours, while the total number of clusters decreases. Here, we only take into account the clusters larger than $82$ particles, since that is the minimum number of particles to form the $32$-particle icosahedral inner shell with the first layer around it containing $50$ particles. Our simulations suggest that the unit cell starts growing from these structures. Given these results, we have chosen the $11$ species and $29$ colours design (Fig.~\ref{fig1}) to directly nucleate the target structure. For direct nucleation simulations, we randomly distribute $N=5680$ particles in a cubic box, with the same species ratio as the unit cell. In Fig.~\ref{fig2}b) we present a cluster that nucleated for $T=0.096$ and $\rho=0.1$ and has grown up to $N=2221$ particles. We represent only the bonds between the particles and the cluster oriented along the pseudo-five-fold symmetry axis, clearly showing the intended symmetry. We have further calculated the diffraction pattern along this axis (Fig.~\ref{fig2}c) which highlights the $10$ peaks corresponding to the pseudo-five-fold symmetry, thus confirming that the correct structure has nucleated.

\subsection{Ideal icosahedral quasicrystal}

\begin{figure}[t]
	\includegraphics{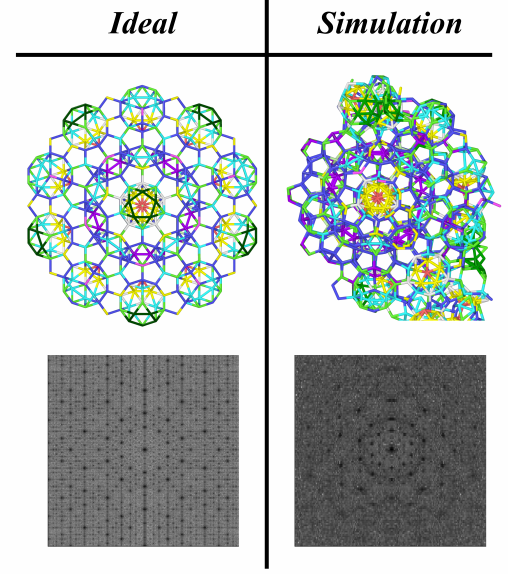}
	\caption{\label{fig3} Direct nucleation of the ideal icosahedral quasicrystal. On the left is shown a portion of the ideal quasicrystal containing $N=1538$ points which is used in SAT to create the patchy particle designs for simulation. On the right is the simulated cluster obtained from direct nucleation simulations with $N=1538$, $T=0.0875$, $\rho=0.05$, and for $\Delta t=10^9$. Both clusters are oriented along the five-fold symmetry axis and the respective diffraction patterns are shown in the bottom panels.}
\end{figure}

Through the cut-and-project method one can project onto physical space the ideal icosahedral quasicrystal by using only irrational cuts on the 6D hypercubic body-centred lattice. Unlike the previous structure, this one is aperiodic and shows five, three and two-fold symmetry. Fig~\ref{fig3} shows a schematic of a portion of the ideal icosahedral quasicrystal ($1538$ particles) where we have connected with lines the bonded particles. We have oriented the structure along the five-fold symmetry axis and present the diffraction pattern along this axis. The total number of unique environments is now $12$ (see \emph{Supporting Information}). These include the ones shown in Fig~\ref{fig1}d) and some subsets of these with fewer patches. For example, one of the new environments corresponds to the $7$-patch yellow particle in Fig~\ref{fig1}d) (first from the left) without one of the patches numbered $15$.

For this structure, we used SAT to calculate patchy particle designs that satisfy the topology of the quasicrystal which includes all the $12$ environments. The minimum number of particles for this condition is around $N=1538$. We still keep the constraint that the designs should not satisfy the topology of the $BC8$ crystal unit cell. In this structure the different environments bond between each other in $28$ unique ways and, thus, we fix the total number of species to this value. By following a similar procedure as in the previous structure one can optimize the design for direct nucleation, which in this case corresponds to a $28$ species and $27$ colours design (details of the design can be found in the \emph{Supporting Information}). We further introduce a two-step protocol for direct nucleation of this structure, where we start the simulations at a higher temperature (around $T=0.12$) to form only the $32$-particle icosahedral inner shell and then instantly quench to a lower temperature ($T=0.0875$) to grow the structure. The icosahedral inner shell is able to form first since it includes the species with the most patches ($7$ and $6$ respectively), leading to a higher bonding probability respective to the others. We have found that forming the initial $32$-particle cluster helps significantly the self-assembly of the rest of the structure. We also note that we were not able to form this structure from the liquid, it only consistently grew at lower densities ($\rho\leq 0.1$) from the gas phase. Fig~\ref{fig3} shows one of the clusters that formed in simulations with $\rho=0.05$ oriented along the five-fold symmetry axis. The diffraction pattern below is calculated along this axis and confirms the assembled structure. Since SAT only uses a subset of the entire quasicrystal, it is not guaranteed that an assembled structure can keep growing after reaching the respective subset size, since it is aperiodic. Thus, to assemble larger structures for the ideal quasicrystal one needs to fully re-SAT a larger subset.

\subsection{Mapping onto icosahedral units}

\begin{figure}[t]
	\includegraphics{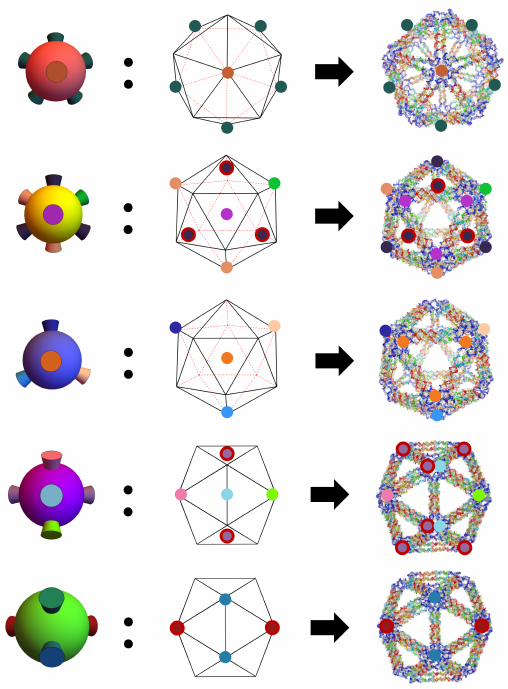}
	\caption{\label{fig4} Schematic representation of the proposed realization of this system through icosahedral DNA origami. Each patchy particle environment corresponds to a given icosahedral DNA origami with patches on selected vertices. For each environment we present a mapping onto the icosahedron and the respective DNA origami. The red dashed lines on the icosahedron correspond to the back faces, and the patch circles highlighted in red are located at the back of the origami, while the others are at the front. For two of the patchy particles used, all patches can be mapped directly to the vertices of the icosahedron (top and bottom environments). For the remainder three, we use a map where a patch that is located on a face of the icosahedron is split into three identical patches at the closest vertices, while if the patch is located on an edge it is split into the two nearest vertices. This mapping guarantees correct alignment of the icosahedral DNA origami, which we previously used for experimental realizations of DNA lattice assemblies \cite{Liu2023}.}
\end{figure}

Recently, patchy particle models have been quite successful as coarse-grained models to study the assembly of DNA origami~\cite{Liu2023}. The quasicrystal designs proposed in the previous section can be mapped onto an icosahedron such that it can be functionalized with single DNA strands at the vertices. Figure~\ref{fig4} shows a possible mapping of the different patch geometries onto an icosahedron DNA origami. In standard DNA origami, single strands of DNA can be functionalized onto the vertices of the wireframe, thus, we have decided to keep the patch positions which are mapped onto the vertices of the icosahedron and redistribute the others that fall on the faces or edges. For this, we have replaced a patch located on the face of the icosahedron by three new patches at the closest vertices, while the ones located at the edges are exchanged by two new patches at the closest vertices. This mapping guarantees proper orientation of the DNA origami wireframe for the assembly of the icosahedral quasicrystal.

\section{Conclusions}

In this work we have proposed a new automated method to generate patchy particle designs with coloured interactions that can self-assemble the icosahedral quasicrystal and its approximants. We use a simple patchy particle model with attractive surface patches which are coloured such that they bond only with the complementary colour. This selective interaction can be described as a SAT problem and efficiently solved numerically to produce designs (patch colour distribution and colour-colour interaction matrix) that target specific structures. One can then test the designs in coarse-grained simulations to understand their efficacy. We demonstrate how to use this strategy to self-assemble the $3/2$ icosahedral quasicrystal approximant, which shows the pseudo-five-fold symmetry axis characteristic of this structure. We then expand this procedure to target the ideal quasicrystal by using a design that satisfies a portion of the structure. Through a 2-step temperature quench protocol, we are able to self-assemble this structure in simulations and highlight its five-fold symmetry axis through the diffraction pattern. Lastly, we introduce a possible mapping of our patchy particle design onto a DNA origami icosahedron. By mapping patches that are positioned at the faces and edges to the closest vertices we are able to propose a DNA origami which only requires DNA strands to be functionalized to the vertices. This guarantees the required orientational order of the icosahedral quasicrystal. Given this mapping and the simplicity of the patchy particle model used, we believe such designs are amenable to experiments as demonstrated for the case of pyrochlore lattice in recent work~\cite{Liu2023}.

Our results demonstrate that using only coloured patch specificity is enough to nucleate these structures. Thus, even with a simple patch-patch interaction the intended orientational symmetry can be propagated. Our method is also fully automated and can be arbitrarily constrained to fit the experimental requirements. For example, for all the designs shown here, each colour only interacts with one other. In the context of DNA origami, our simplification of the patchy particle interactions should allow for an easier translation of the simulations into experiments.

During submission of this manuscript we became aware of the recent work by Noya and Doye~\cite{Noya2024} who rationally designed a one-component patchy particle to self-assemble a face-centered icosahedral quasicrystal. The advantage of their solution is the simplicity of the building block, but it requires two-energy scales. Our method instead suggests the use of multiple species in order to work with a single energy scale and additional experimental constraints, such as having orthogonal interactions (i.e. each colour has only one complementary colour and does not interact with other colours) which can be translated into DNA-sequences more easily.

\section{Methods}

We consider a system composed of $N$ patchy particles in a cubic box of length $L$. Particles are characterized by a hard core of diameter $\sigma$ with $N_{patches}$ patches on its surface. The patches interact through the Kern-Frenkel potential \citep{Kern2003, bol1982monte}:

\begin{equation}
Vpp(\boldsymbol{r}_{ij}, \boldsymbol{\hat{r}}_{\alpha, i}, \boldsymbol{\hat{r}}_{\beta, j})=V_{SW}(r_{ij})f(\boldsymbol{r}_{ij}, \boldsymbol{\hat{r}}_{\alpha, i}, \boldsymbol{\hat{r}}_{\beta, j})
\end{equation}

where $i$ corresponds to a given particle and $\boldsymbol{r}_{i}$ its center of mass. Thus, $\boldsymbol{r}_{ij}$ is the distance between particles $i$ and $j$. $\boldsymbol{r}_{\alpha, i}$ denotes the position of patch $\alpha$ of particle $i$. $V_{SW}$ is an isotropic square-well of range $\sigma + \delta_{\alpha,\beta}$ and depth $\varepsilon_{\alpha,\beta}$, the hat symbol indicate unit vectors and $f$ is the orientation-dependent modulation term that takes the form:

\begin{equation}
\label{KF}
f(\boldsymbol{r}_{ij}, \boldsymbol{\hat{r}}_{\alpha, i}, \boldsymbol{\hat{r}}_{\beta, j})=
\begin{cases}
1 & \text{if $\begin{aligned}
	\text{$\boldsymbol{\hat{r}}_{ij} \cdot \boldsymbol{\hat{r}}_{\alpha, i} > \cos     \theta^{max}_{\alpha \beta}$} \\
	\text{$\boldsymbol{\hat{r}}_{ji} \cdot \boldsymbol{\hat{r}}_{\beta, j} > \cos     \theta^{max}_{\alpha \beta}$}
	\end{aligned}$ } \\
0 & \text{otherwise}
\end{cases}
\end{equation}

With this formulation patches are represented by a cone starting from the center of mass of the particle and reaching $\sigma + \delta_{\alpha,\beta}$, while the width is controlled by $\theta^{max}_{\alpha \beta}$. For simplicity, we consider the parameter range where it is only possible to form one bond per patch. In the following, $\sigma$ provides the unit of length and $\varepsilon_{\alpha, \beta}$ the unit of energy. Temperature ($T$) is also expressed in units of $\varepsilon_{\alpha, \beta}$ and $k_B=1$.

For the results we considered Monte Carlo (MC) simulations with two possible moves, roto-translations and aggregation-volume-bias~\citep{Rovigatti2018}. The first attempts a simple rotation and translation of a random particle along a random (radial or angular) direction. The second, attempts to move a random particle into the vicinity of another such that a bond is formed between the two. To not break ergodicity, the inverse move can also be performed where a random bond between two particles is broken. We performed simulations in the $NVT$ assemble to explore the assembly of the desired structures. For the results shown, we considered $\delta_{\alpha, \beta}=0.2$ and $\cos\theta^{max}_{\alpha \beta}=0.98$. The direct nucleation simulations start with particles randomly generated in the box with random orientations, while the seeding simulations start with the seed placed at the center of the box and the rest of the particles randomly distributed, with a buffer space between the seed and the rest of the particles with size of the order of a single particle diameter.

The code implementing the SAT-assembly pipeline is available at https://github.com/deppinto/PatchyParticles.

\section{Acknowledgments}

We acknowledge partial support by ICSC -- Centro Nazionale di Ricerca in High Performance Computing, Big Data and Quantum Computing, funded by European Union -- NextGenerationEU, and the CINECA award under the ISCRA initiative, for the availability of high-performance computing resources and support. D.P. and  F.S. acknowledge  support from MIUR-PRIN Grant No. 2022JWAF7Y. We thank Michael Matthies for help with visualization of DNA origami.

\section{Supporting Information}

\subsection{SAT framework for the $3/2$ icosahedral quasicrystal approximant}

For our assembly method, we need to translate the patchy particle design into a SAT problem of boolean variables and then impose the relevant constraints between these variables to form the target structure. In our framework we use a super-particle description which merges all the different environments into one single particle, then, through proper distribution of patch colours and colour-colour interactions we can selectively activate only a subset $N_{subset}$ of the total number of patches $N_{patches}$ for each particle. In Fig~\ref{fig1}, only the active patches are shown, since the rest are given a colour that does not have a complementary interaction, thus are inert. In Fig~\ref{fig1}d) we show schematically the different environments that form the super-particles used. The positions of the $N_{patches}=31$ super-particle patches are given as:

\begin{multline}
\\
\textbf{p}_1=(0.58, 0.58, 0.58) \\
\textbf{p}_2=(-0.85, 0.53, 0) \\
\textbf{p}_3=(0.53, 0, -0.85) \\
\textbf{p}_4=(0, -0.85, 0.53) \\
\textbf{p}_5=(-0.85, 0.53, 0) \\
\textbf{p}_6=(0.85, 0.53, 0) \\
\textbf{p}_7=(0, 1, 0) \\
\textbf{p}_8=(0, -0.36, 0.93) \\
\textbf{p}_9=(0, -0.36, -0.93) \\
\textbf{p}_{10}=(0.85, -0.53, 0) \\
\textbf{p}_{11}=(0.85, 0.53, 0) \\
\textbf{p}_{12}=(-0.53, 0, -0.85) \\
\textbf{p}_{13}=(-0.53, 0, 0.85) \\
\textbf{p}_{14}=(-0.85, 0.53, 0) \\
\textbf{p}_{15}=(0.85, 0.53, 0) \\
\textbf{p}_{16}=(0, -0.85, 0.53) \\
\textbf{p}_{17}=(0.53, 0, -0.85) \\
\textbf{p}_{18}=(0.53, 0, 0.85) \\
\textbf{p}_{19}=(0, -0.36, -0.93) \\
\textbf{p}_{20}=(-0.81, -0.31, 0.5) \\
\textbf{p}_{21}=(0.81, -0.31, 0.5) \\
\textbf{p}_{22}=(0, 1, 0) \\
\textbf{p}_{23}=(0, -0.85, 0.53) \\
\textbf{p}_{24}=(0.85, 0.53, -0) \\
\textbf{p}_{25}=(-0.85, 0.53, -0) \\
\textbf{p}_{26}=(-0.53, 0, -0.85) \\
\textbf{p}_{27}=(0.85, -0.53, -0) \\
\textbf{p}_{28}=(0.85, 0.53, -0) \\
\textbf{p}_{29}=(0, 0.85, 0.53) \\
\textbf{p}_{30}=(-0.53, 0, 0.85) \\
\textbf{p}_{31}=(0, -0.85, 0.53) \\
\end{multline}

Where they can be distributed by the coloured environments in Fig.~\ref{fig1} in the following way: 

\begin{multline}
\\
Blue = (p_{1}, p_{2}, p_{3}, p_{4}) \\
Purple = (p_{5}, p_{6}, p_{7}, p_{8}, p_{9}) \\
Green = (p_{10}, p_{11}, p_{12}, p_{13}) \\
Brown = (p_{14}, p_{15}, p_{16}) \\
Magenta = (p_{17}, p_{18}) \\
Yellow = (p_{19}, p_{20}, p_{21}, p_{22}, p_{23}, p_{24}, p_{25}) \\
Red = (p_{26}, p_{27}, p_{28}, p_{29}, p_{30}, p_{31}) \\
\end{multline}

The numbered patch colours shown in Fig.~\ref{fig1} can be similarly ordered in the following way:

\begin{multline}
\\
1-Blue = (20, 21, 8, 14) \\
2-Blue = (19, 23, 23, 0) \\
3-Blue = (25, 26, 11, 27) \\
4-Blue = (1, 18, 21, 14) \\
5-Blue = (20, 7, 7, 7) \\
Purple = (24, 2, 5, 28, 28) \\
Green = (10, 4, 13, 13) \\
Brown = (17, 26, 27) \\
Magenta = (12 ,12) \\
Yellow = (22, 15, 3, 15, 9, 9, 9) \\
Red = (16, 6, 6, 6, 6, 6) \\
\end{multline}

Lastly, the ratio between the particle species is the following:

\begin{multline}
\\
1-Blue = 16/284 \\
2-Blue =  48/284 \\
3-Blue =  48/284 \\
4-Blue =  24/284 \\
5-Blue =  24/284 \\
Purple =  12/284 \\
Green =  12/284 \\
Brown =  24/284 \\
Magenta =  12/284 \\
Yellow =  40/284 \\
Red =  24/284 \\
\end{multline}

The boolean variables can be organized into four major groups. The first corresponds to the colour interaction variables, $x_{c_i,c_j}^{int}$, where $c_i$ and $c_j$ are the colour of particle $i$ and $j$ respectively. If this variable is true then colours $c_i$ and $c_j$ interact and can form a bond, otherwise not. There are a total of $(N_c)(N_c+1)/2$ of these variables. The second group is the patch colouring variable, $x_{p,s,c}^{pcol}$, where $p\in [1, N_p]$ refers to particle species, $s\in [1, V]$ to patch number and colour $c\in [1, N_c]$. If true, particle specie $p$ has the patch number $s$ of colour $c$. There are $N_pVN_c$ of these variables. Then the placement variables, $x_{l,p,o}^{L}$, where $l\in [1, L]$ refers the position of a particle in the target structure, $p\in [1, N_p]$ to particle specie and orientation $o\in [1, R]$. If true, a particle of species $p$ occupies position $l$ according to orientation $o$. There are $N_pLR$ of these variables. Lastly, there is an auxiliary variable, $x_{l,s,c}^{A}$. If true, the particle in position $l$ is oriented such that the patch $s$ has a colour $c$. There are $VLN_c$ such variables. The orientation mapping is given in Table I.

There are seven main groups of clauses solved by SAT. The first guarantees that each colour can only interact with only one other colour:

\begin{equation}
C^{int}_{c_i,c_j,c_k}=\neg x_{c_i,c_j}^{int} \vee \neg x_{c_i,c_k}^{int}
\end{equation}

The second ensures that patch number $s$ of particle specie $p$ will have exactly one colour only:

\begin{equation}
C^{pcol}_{p,s,c_k,c_l}=\neg x_{p,s,c_k}^{pcol} \vee \neg x_{p,s,c_l}^{pcol}
\end{equation}

The third guarantees that position $l$ is occupied by exactly one particle specie with one orientation:

\begin{equation}
C^{L}_{l,p_i,o_i,p_j,o_j}=\neg x_{l,p_i,o_i}^{L} \vee \neg x_{l,p_j,o_j}^{L}
\end{equation}

The fourth enforces that the neighbouring positions $l_i$ and $l_j$ connected by the patches $s_i$ and $s_j$ have colours in those patches, $c_i$ and $c_j$, which interact:

\begin{equation}
C^{lint}_{l_i,s_i,l_j,s_j,c_i,c_j}=\neg x_{l_i,s_i,c_i}^{A} \vee \neg x_{l_j,s_j,c_j}^{A} \vee x_{c_i,c_j}^{int}
\end{equation}

The fifth ensures that for a position $l$ that is occupied by particle specie $p$ with orientation $o$, the patch $s$ has the right colour attributed to it:

\begin{equation}
\begin{split}
C^{LS}_{l,p,o,c,s}= & ( \neg x_{l,p,o}^{L} \vee \neg x_{l,s,c}^{A} \vee x_{p,\phi_o(s), c}^{pcol} ) \\ 
& \wedge ( \neg x_{l,p,o}^{L} \vee x_{l,s,c}^{A} \vee \neg x_{p,\phi_o(s), c}^{pcol} )
\end{split}
\end{equation}

The two last groups define multiple clauses each, the first enforces that all particle species are used, the second enforces that all colours are used to form the topology off the structure:

\begin{equation}
\forall p\in [1, N_p]:C_p^{all p.}= \underset{\forall l \in [1,L], o\in [1,R]}{\bigvee} x_{l,p,o}^L
\end{equation}

\begin{equation}
\forall c\in [1, N_c]:C_c^{all c.}= \underset{\forall l \in [1,L], s\in [1,V]}{\bigvee} x_{l,s,c}^{A}
\end{equation}

To optimize the SAT solver and reduce the total number of clauses, we also fix the species of every particle in the structure. For that we calculate the first neighbours of every particle $i$ and see to which environments each pair neighbour belongs to. For every unique (cannot be obtained through rigid body rotations) pair of environments (particle $i$ plus neighbour) we add one more specie of the particle $i$ environment. By creating this map between particle position and specie we can simplify all the above clauses to only run through the singular specie that a position belongs to.

\begin{table*}
	\begin{center}
		\begin{tabular}{ cccc } 
			\hline
			Orientation $o$ & Mapping to patch number \\
			\hline
1 & (17:17 ,18:18 ,14:14 ,15:15 ,16:16 ,1:1 ,2:2 ,3:3 ,4:4 ,10:10 ,11:11 ,12:12 ,13:13 ,5:5 ,6:6 ,7:7 , \\
& 8:8 ,9:9 ,26:26 ,27:27 ,28:28 ,29:29 ,30:30 ,31:31 ,19:19 ,20:20 ,21:21 ,22:22 ,23:23 ,24:24 ,25:25) \\

2 & (17:17 ,18:18 ,14:14 ,15:15 ,16:16 ,1:1 ,2:4 ,3:2 ,4:3 ,10:10 ,11:11 ,12:12 ,13:13 ,5:5 ,6:6 ,7:7 , \\
& 8:8 ,9:9 ,26:26 ,27:27 ,28:28 ,29:29 ,30:30 ,31:31 ,19:19 ,20:20 ,21:21 ,22:22 ,23:23 ,24:24 ,25:25) \\

3 & (17:17 ,18:18 ,14:14 ,15:15 ,16:16 ,1:1 ,2:3 ,3:4 ,4:2 ,10:10 ,11:11 ,12:12 ,13:13 ,5:5 ,6:6 ,7:7 , \\
& 8:8 ,9:9 ,26:26 ,27:27 ,28:28 ,29:29 ,30:30 ,31:31 ,19:19 ,20:20 ,21:21 ,22:22 ,23:23 ,24:24 ,25:25) \\

4 & (17:17 ,18:18 ,14:14 ,15:15 ,16:16 ,1:1 ,2:2 ,3:3 ,4:4 ,10:10 ,11:11 ,12:12 ,13:13 ,5:6 ,6:5 ,7:7 , \\
& 8:9 ,9:8 ,26:26 ,27:27 ,28:28 ,29:29 ,30:30 ,31:31 ,19:19 ,20:20 ,21:21 ,22:22 ,23:23 ,24:24 ,25:25) \\

5 & (17:17 ,18:18 ,14:14 ,15:15 ,16:16 ,1:1 ,2:2 ,3:3 ,4:4 ,10:11 ,11:10 ,12:13 ,13:12 ,5:5 ,6:6 ,7:7 , \\
& 8:8 ,9:9 ,26:26 ,27:27 ,28:28 ,29:29 ,30:30 ,31:31 ,19:19 ,20:20 ,21:21 ,22:22 ,23:23 ,24:24 ,25:25) \\

6 & (17:17 ,18:18 ,14:16 ,15:14 ,16:15 ,1:1 ,2:2 ,3:3 ,4:4 ,10:10 ,11:11 ,12:12 ,13:13 ,5:5 ,6:6 ,7:7 , \\
& 8:8 ,9:9 ,26:26 ,27:27 ,28:28 ,29:29 ,30:30 ,31:31 ,19:19 ,20:20 ,21:21 ,22:22 ,23:23 ,24:24 ,25:25) \\

7 & (17:17 ,18:18 ,14:15 ,15:16 ,16:14 ,1:1 ,2:2 ,3:3 ,4:4 ,10:10 ,11:11 ,12:12 ,13:13 ,5:5 ,6:6 ,7:7 , \\
& 8:8 ,9:9 ,26:26 ,27:27 ,28:28 ,29:29 ,30:30 ,31:31 ,19:19 ,20:20 ,21:21 ,22:22 ,23:23 ,24:24 ,25:25) \\

8 & (17:18 ,18:17 ,14:14 ,15:15 ,16:16 ,1:1 ,2:2 ,3:3 ,4:4 ,10:10 ,11:11 ,12:12 ,13:13 ,5:5 ,6:6 ,7:7 , \\ 
& 8:8 ,9:9 ,26:26 ,27:27 ,28:28 ,29:29 ,30:30 ,31:31 ,19:19 ,20:20 ,21:21 ,22:22 ,23:23 ,24:24 ,25:25) \\

9 & (17:17 ,18:18 ,14:14 ,15:15 ,16:16 ,1:1 ,2:2 ,3:3 ,4:4 ,10:10 ,11:11 ,12:12 ,13:13 ,5:5 ,6:6 ,7:7 , \\
& 8:8 ,9:9 ,26:26 ,27:27 ,28:28 ,29:29 ,30:30 ,31:31 ,19:19 ,20:22 ,21:20 ,22:21 ,23:25 ,24:23 ,25:24) \\

10 & (17:17 ,18:18 ,14:14 ,15:15 ,16:16 ,1:1 ,2:2 ,3:3 ,4:4 ,10:10 ,11:11 ,12:12 ,13:13 ,5:5 ,6:6 ,7:7 , \\
& 8:8 ,9:9 ,26:26 ,27:27 ,28:28 ,29:29 ,30:30 ,31:31 ,19:19 ,20:21 ,21:22 ,22:20 ,23:24 ,24:25 ,25:23) \\

11 & (17:17 ,18:18 ,14:14 ,15:15 ,16:16 ,1:1 ,2:2 ,3:3 ,4:4 ,10:10 ,11:11 ,12:12 ,13:13 ,5:5 ,6:6 ,7:7 , \\
& 8:8 ,9:9 ,26:26 ,27:31 ,28:27 ,29:28 ,30:29 ,31:30 ,19:19 ,20:20 ,21:21 ,22:22 ,23:23 ,24:24 ,25:25) \\

12 & (17:17 ,18:18 ,14:14 ,15:15 ,16:16 ,1:1 ,2:2 ,3:3 ,4:4 ,10:10 ,11:11 ,12:12 ,13:13 ,5:5 ,6:6 ,7:7 , \\
& 8:8 ,9:9 ,26:26 ,27:30 ,28:31 ,29:27 ,30:28 ,31:29 ,19:19 ,20:20 ,21:21 ,22:22 ,23:23 ,24:24 ,25:25) \\

13 & (17:17 ,18:18 ,14:14 ,15:15 ,16:16 ,1:1 ,2:2 ,3:3 ,4:4 ,10:10 ,11:11 ,12:12 ,13:13 ,5:5 ,6:6 ,7:7 , \\
& 8:8 ,9:9 ,26:26 ,27:29 ,28:30 ,29:31 ,30:27 ,31:28 ,19:19 ,20:20 ,21:21 ,22:22 ,23:23 ,24:24 ,25:25) \\

14 & (17:17 ,18:18 ,14:14 ,15:15 ,16:16 ,1:1 ,2:2 ,3:3 ,4:4 ,10:10 ,11:11 ,12:12 ,13:13 ,5:5 ,6:6 ,7:7 , \\
& 8:8 ,9:9 ,26:26 ,27:28 ,28:29 ,29:30 ,30:31 ,31:27 ,19:19 ,20:20 ,21:21 ,22:22 ,23:23 ,24:24 ,25:25) \\

			\hline
		\end{tabular}
		\caption{Mapping of the orientation to the patch numbers for the $3/2$ icosahedral quasicrystal approximant.}
	\end{center}
\end{table*}

\subsection{SAT framework for the ideal icosahedral quasicrystal}

We again use a super-particle description which now encapsulates $12$ particle environments as patch geometries. Then, we use the patch colouring and colour interaction matrix to allow only a subset of the patches to interact depending on the super-particle specie. The positions of the patches are shown in Table II.

\begin{table*}
    \begin{center}
    \begin{tabular}{ cccc } 
    \hline
$\textbf{p}_{0}=(0.58,-0.19,-0.79)$ & $\textbf{p}_{1}=(0.53,-0.45,0.72)$ \\
$\textbf{p}_{2}=(-0.85,-0.45,-0.28)$ & $\textbf{p}_{3}=(0,1,0)$ \\
$\textbf{p}_{4}=(0,0.45,0.89)$ & $\textbf{p}_{4}=(0,-1,0)$ \\
$\textbf{p}_{6}=(0.53,0.45,-0.72)$ & $\textbf{p}_{7}=(-0.53,-0.45,0.72)$ \\
$\textbf{p}_{8}=(0.85,-0.45,-0.28)$ & $\textbf{p}_{9}=(0.31,-0.85,0.43)$ \\
$\textbf{p}_{10}=(0.36,0.79,0.49)$ & $\textbf{p}_{11}=(-0.58,-0.19,-0.79)$ \\
$\textbf{p}_{12}=(0.53,0.45,-0.72)$ & $\textbf{p}_{13}=(0.85,-0.45,-0.28)$ \\
$\textbf{p}_{14}=(-0.85,-0.45,-0.28)$ & $\textbf{p}_{15}=(0,0.45,0.89)$ \\
$\textbf{p}_{16}=(-0.93,-0.19,0.3)$ & $\textbf{p}_{17}=(0.58,-0.19,-0.79)$ \\
$\textbf{p}_{18}=(0.53,-0.45,0.72)$ & $\textbf{p}_{19}=(0,1,0)$ \\
$\textbf{p}_{20}=(0,-0.45,-0.89)$ & $\textbf{p}_{21}=(0,-0.19,0.98)$ \\
$\textbf{p}_{22}=(0.81,0.53,-0.26)$ & $\textbf{p}_{23}=(-0.81,0.53,-0.26)$ \\
$\textbf{p}_{24}=(0,-0.85,-0.53)$ & $\textbf{p}_{25}=(0,1,0)$ \\
$\textbf{p}_{26}=(-0.85,-0.45,-0.28)$ & $\textbf{p}_{27}=(0.85,-0.45,-0.28)$ \\
$\textbf{p}_{28}=(0.53,-0.45,0.72)$ & $\textbf{p}_{29}=(-0.85,0.45,0.28)$ \\
$\textbf{p}_{30}=(-0.85,-0.45,-0.28)$ & $\textbf{p}_{31}=(0,-0.45,-0.89)$ \\
$\textbf{p}_{32}=(0.53,0.45,-0.72)$ & $\textbf{p}_{33}=(0,1,0)$ \\
$\textbf{p}_{34}=(0,-0.19,0.98)$ & $\textbf{p}_{35}=(0.85,-0.45,-0.28)$ \\
$\textbf{p}_{36}=(0.81,0.53,-0.26)$ & $\textbf{p}_{37}=(0,1,0)$ \\
$\textbf{p}_{38}=(-0.81,0.53,-0.26)$ & $\textbf{p}_{39}=(-0.85,-0.45,-0.28)$ \\
$\textbf{p}_{40}=(-0.36,0.79,0.49)$ & $\textbf{p}_{41}=(-0.31,-0.85,0.43)$ \\
$\textbf{p}_{42}=(0.53,-0.45,0.72)$ & $\textbf{p}_{43}=(-0.85,-0.45,-0.28)$ \\
$\textbf{p}_{44}=(0.53,0.45,-0.72)$ & $\textbf{p}_{45}=(0,-1,0)$ \\
$\textbf{p}_{46}=(-0.53,0.45,-0.72)$ & $\textbf{p}_{47}=(0.85,-0.45,-0.28)$ \\
$\textbf{p}_{48}=(-0.53,-0.45,0.72)$ & $\textbf{p}_{49}=(-0.85,-0.45,-0.28)$ \\
$\textbf{p}_{50}=(-0.53,0.45,-0.72)$ & $\textbf{p}_{51}=(0,1,0)$ \\
        \hline
        \end{tabular}
        \caption{Patch position for the ideal icosahedral quasicrystal.}
    \end{center}
\end{table*}

The clauses for SAT are the same as the ones in the approximant, only the target topology changes. The orientation mapping is given in Table III.

\begin{table*}
	\begin{center}
		\begin{tabular}{ cccc } 
			\hline
			Orientation $o$ & Mapping to patch number \\
			\hline
1 & (20:20,45:45,46:46,4:4,5:5,6:6,0:0,1:1,2:2,3:3,12:12,13:13,14:14, 15:15, 16:16, \\
 & 17:17,18:18, 19:19, 7:7,8:8,9:9, 10:10,11:11, 40:40,41:41,42:42, 43:43, 44:44, \\
 & 47:47,48:48,49:49,50:50,51:51,28:28,29:29, 30:30,31:31,32:32, 33:33, 34:34, \\
 & 35:35,36:36,37:37,38:38,39:39,21:21,22:22,23:23,24:24,25:25,26:26,27:27) \\
 
 2 & (20:20,45:45,46:46,4:4,5:5,6:6,0:0,1:3,2:1,3:2,12:12,13:13,14:14,15:15,16:16, \\
 & 17:17,18:18,19:19,7:7,8:8,9:9,10:10,11:11,40:40,41:41,42:42,43:43,44:44, \\
 & 47:47,48:48,49:49,50:50,51:51,28:28,29:29,30:30,31:31,32:32,33:33,34:34, \\
 & 35:35,36:36,37:37,38:38,39:39,21:21,22:22,23:23,24:24,25:25,26:26,27:27) \\
 
 3 & (20:20,45:45,46:46,4:4,5:5,6:6,0:0,1:2,2:3,3:1,12:12,13:13,14:14,15:15,16:16, \\
 & 17:17,18:18,19:19,7:7,8:8,9:9,10:10,11:11,40:40,41:41,42:42,43:43,44:44, \\
 & 47:47,48:48,49:49,50:50,51:51,28:28,29:29,30:30,31:31,32:32,33:33,34:34, \\
 & 35:35,36:36,37:37,38:38,39:39,21:21,22:22,23:23,24:24,25:25,26:26,27:27) \\
 
4 & (20:20,45:45,46:46,4:6,5:4,6:5,0:0,1:1,2:2,3:3,12:12,13:13,14:14,15:15,16:16, \\
& 17:17,18:18,19:19,7:7,8:8,9:9,10:10,11:11,40:40,41:41,42:42,43:43,44:44, \\ 
& 47:47,48:48,49:49,50:50,51:51,28:28,29:29,30:30,31:31,32:32,33:33,34:34, \\
& 35:35,36:36,37:37,38:38,39:39,21:21,22:22,23:23,24:24,25:25,26:26,27:27) \\

5 & (20:20,45:45,46:46,4:5,5:6,6:4,0:0,1:1,2:2,3:3,12:12,13:13,14:14,15:15,16:16, \\
& 17:17,18:18,19:19,7:7,8:8,9:9,10:10,11:11,40:40,41:41,42:42,43:43,44:44, \\
& 47:47,48:48,49:49,50:50,51:51,28:28,29:29,30:30,31:31,32:32,33:33,34:34, \\
& 35:35,36:36,37:37,38:38,39:39,21:21,22:22,23:23,24:24,25:25,26:26,27:27) \\

6 & (20:20,45:45,46:46,4:4,5:5,6:6,0:0,1:1,2:2,3:3,12:12,13:13,14:14,15:15,16:16, \\
& 17:17,18:18,19:19,7:8,8:7,9:9,10:11,11:10,40:40,41:41,42:42,43:43,44:44, \\
& 47:47,48:48,49:49,50:50,51:51,28:28,29:29,30:30,31:31,32:32,33:33,34:34, \\
& 35:35,36:36,37:37,38:38,39:39,21:21,22:22,23:23,24:24,25:25,26:26,27:27) \\

7 & (20:20,45:45,46:46,4:4,5:5,6:6,0:0,1:1,2:2,3:3,12:13,13:12,14:15,15:14,16:16, \\
& 17:17,18:18,19:19,7:7,8:8,9:9,10:10,11:11,40:40,41:41,42:42,43:43,44:44, \\
& 47:47,48:48,49:49,50:50,51:51,28:28,29:29,30:30,31:31,32:32,33:33,34:34, \\
& 35:35,36:36,37:37,38:38,39:39,21:21,22:22,23:23,24:24,25:25,26:26,27:27) \\

8 & (20:20,45:45,46:46,4:4,5:5,6:6,0:0,1:1,2:2,3:3,12:12,13:13,14:14,15:15,16:17, \\
& 17:16,18:19,19:18,7:7,8:8,9:9,10:10,11:11,40:40,41:41,42:42,43:43,44:44, \\
& 47:47,48:48,49:49,50:50,51:51,28:28,29:29,30:30,31:31,32:32,33:33,34:34, \\
& 35:35,36:36,37:37,38:38,39:39,21:21,22:22,23:23,24:24,25:25,26:26,27:27) \\

9 & (20:20,45:45,46:46,4:4,5:5,6:6,0:0,1:1,2:2,3:3,12:12,13:13,14:14,15:15,16:16, \\
& 17:17,18:18,19:19,7:7,8:8,9:9,10:10,11:11,40:40,41:41,42:42,43:43,44:44, \\
& 47:47,48:48,49:49,50:50,51:51,28:28,29:29,30:30,31:31,32:32,33:33,34:34, \\
& 35:35,36:36,37:37,38:38,39:39,21:21,22:24,23:22,24:23,25:27,26:25,27:26) \\

10 & (20:20,45:45,46:46,4:4,5:5,6:6,0:0,1:1,2:2,3:3,12:12,13:13,14:14,15:15,16:16, \\
& 17:17,18:18,19:19,7:7,8:8,9:9,10:10,11:11,40:40,41:41,42:42,43:43,44:44, \\
& 47:47,48:48,49:49,50:50,51:51,28:28,29:29,30:30,31:31,32:32,33:33,34:34, \\
& 35:35,36:36,37:37,38:38,39:39,21:21,22:23,23:24,24:22,25:26,26:27,27:25) \\

11 & (20:20,45:45,46:46,4:4,5:5,6:6,0:0,1:1,2:2,3:3,12:12,13:13,14:14,15:15,16:16, \\
& 17:17,18:18,19:19,7:7,8:8,9:9,10:10,11:11,40:40,41:41,42:42,43:43,44:44, \\
& 47:47,48:48,49:49,50:50,51:51,28:28,29:33,30:29,31:30,32:31,33:32,34:34, \\
& 35:35,36:36,37:37,38:38,39:39,21:21,22:22,23:23,24:24,25:25,26:26,27:27) \\

12 & (20:20,45:45,46:46,4:4,5:5,6:6,0:0,1:1,2:2,3:3,12:12,13:13,14:14,15:15,16:16, \\
& 17:17,18:18,19:19,7:7,8:8,9:9,10:10,11:11,40:40,41:41,42:42,43:43,44:44, \\
& 47:47,48:48,49:49,50:50,51:51,28:28,29:32,30:33,31:29,32:30,33:31,34:34, \\
& 35:35,36:36,37:37,38:38,39:39,21:21,22:22,23:23,24:24,25:25,26:26,27:27) \\

13 & (20:20,45:45,46:46,4:4,5:5,6:6,0:0,1:1,2:2,3:3,12:12,13:13,14:14,15:15,16:16, \\
& 17:17,18:18,19:19,7:7,8:8,9:9,10:10,11:11,40:40,41:41,42:42,43:43,44:44, \\
& 47:47,48:48,49:49,50:50,51:51,28:28,29:31,30:32,31:33,32:29,33:30,34:34, \\
& 35:35,36:36,37:37,38:38,39:39,21:21,22:22,23:23,24:24,25:25,26:26,27:27) \\

14 & (20:20,45:45,46:46,4:4,5:5,6:6,0:0,1:1,2:2,3:3,12:12,13:13,14:14,15:15,16:16, \\
& 17:17,18:18,19:19,7:7,8:8,9:9,10:10,11:11,40:40,41:41,42:42,43:43,44:44, \\
& 47:47,48:48,49:49,50:50,51:51,28:28,29:30,30:31,31:32,32:33,33:29,34:34, \\
& 35:35,36:36,37:37,38:38,39:39,21:21,22:22,23:23,24:24,25:25,26:26,27:27) \\

15 & (20:20,45:46,46:45,4:4,5:5,6:6,0:0,1:1,2:2,3:3,12:12,13:13,14:14,15:15,16:16, \\
& 17:17,18:18,19:19,7:7,8:8,9:9,10:10,11:11,40:40,41:41,42:42,43:43,44:44, \\
& 47:47,48:48,49:49,50:50,51:51,28:28,29:29,30:30,31:31,32:32,33:33,34:34, \\
& 35:35,36:36,37:37,38:38,39:39,21:21,22:22,23:23,24:24,25:25,26:26,27:27) \\
			\hline
		\end{tabular}
		\caption{Mapping of the orientation to the patch numbers for the ideal icosahedral quasicrystal.}
	\end{center}
\end{table*}

The design used for the ideal quasicrystal is shown in Table IV below with the following format: $B(X,Y)$ represents the interaction between two colours $X$ and $Y$; $C(X,Y,Z)$ represents the patch colour of each specie, where $X$ gives the specie, $Y$ the patch number and $Z$ the patch colour. We only show the active patches per specie, the remaining are inert.

\begin{table*}
    \begin{center}
    \begin{tabular}{ cccccccc } 
    \hline
B(0,0) & B(1,18) & B(2,15) & B(3,16) & B(4,7) & B(5,24) & B(6,11) & B(8,19) \\
B(9,9) & B(10,13) & B(12,12) & B(14,26) & B(17,22) & B(20,25) & B(21,23) &  \\
\hline
C(1,0,18) & C(1,1,22) & C(1,2,18) & C(1,3,22) & C(2,0,7) & C(2,1,6) & C(2,2,8) & C(2,3,6) \\
C(3,0,11) & C(3,1,11) & C(3,2,23) & C(3,3,19) & C(4,0,10) & C(4,1,23) & C(4,2,23) & C(4,3,23) \\
C(5,0,0) & C(5,1,15) & C(5,2,15) & C(5,3,15) & C(6,0,11) & C(6,1,24) & C(6,2,18) & C(6,3,24) \\
C(7,0,18) & C(7,1,18) & C(7,2,18) & C(7,3,25) & C(8,0,4) & C(8,1,22) & C(8,2,16) & C(8,3,8) \\
C(9,0,24) & C(9,1,19) & C(9,2,23) & C(9,3,17) & C(10,0,4) & C(10,1,17) & C(10,2,23) & C(10,3,20) \\
C(11,4,1) & C(11,5,17) & C(11,6,1) & C(12,4,0) & C(12,5,0) & C(12,6,0) & C(13,7,14) & C(13,8,14) \\
C(13,9,9) & C(13,10,8) & C(13,11,8) & C(14,7,14) & C(14,8,14) & C(14,9,12) & C(14,10,6) & C(14,11,6) \\
C(15,12,19) & C(15,13,19) & C(15,14,11) & C(15,15,20) & C(16,12,26) & C(16,13,26) & C(16,14,21) & C(16,15,21) \\
C(17,12,19) & C(17,13,19) & C(17,14,2) & C(17,15,2) & C(18,12,2) & C(18,13,2) & C(18,14,1) & 
C(18,15,1)  \\ 
C(19,16,19) & C(19,17,19) & C(19,18,0) & C(19,19,3) & C(20,20,4) & C(21,21,5) & C(21,22,6) & C(21,23,8)  \\
C(21,24,19) & C(21,25,17) & C(21,26,17) & C(21,27,17) & C(22,21,13) & C(22,22,0) & C(22,23,0) & C(22,24,0) \\
C(22,25,17) & C(22,26,17) & C(22,27,17) & C(23,28,7) & C(23,29,22) & C(23,30,22) & C(23,31,22) & C(23,32,22) \\
C(23,33,22) & C(24,34,0) & C(24,35,15) & C(24,36,6) & C(24,37,0) & C(24,38,11) & C(24,39,15) & C(25,40,1) \\
C(25,41,12) & C(25,42,8) & C(25,43,8) & C(25,44,5) & C(26,40,7) & C(26,41,11) & C(26,42,10) & C(26,43,20) \\
C(26,44,18) & C(27,45,1) & C(27,46,1) & C(28,47,25) & C(28,48,13) & C(28,49,22) & C(28,50,22) & C(28,51,25) \\
        \hline
        \end{tabular}
        \caption{Design for the ideal quasicrystal.}
    \end{center}
\end{table*}

The ratio between the different species is the following:

\begin{multline}
\\
S_{1}= 60/1538, S_{2}= 60/1538 \\
S_{3}= 60/1538, S_{4}= 20/1538 \\
S_{5}= 20/1538, S_{6}= 60/1538 \\
S_{7}= 60/1538, S_{8}= 60/1538 \\
S_{9}= 60/1538, S_{10}= 60/1538 \\
S_{11}= 60/1538, S_{12}= 20/1538 \\
S_{13}= 60/1538, S_{14}= 60/1538 \\
S_{15}= 60/1538, S_{16}= 120/1538 \\
S_{17}= 60/1538, S_{18}= 60/1538 \\
S_{19}= 60/1538, S_{20}= 24/1538 \\
S_{21}= 60/1538, S_{22}= 20/1538 \\
S_{23}= 24/1538, S_{24}= 60/1538 \\
S_{25}= 120/1538, S_{26}= 60/1538 \\
S_{27}= 30/1538, S_{28}= 60/1538 \\
\end{multline}

\subsection{The cut-and-project method}

We use the cut-and-project method to generate the ideal icosahedral quasicrystal and its approximant. This method consists on projecting a quasicrystal from a higher dimensional lattice that has been cut, thus, only a subset of points are projected. For the ideal icosahedral quasicrystal we use a 6D hypercubic body-centred lattice. The 6D space can be divided into two subspaces, the perpendicular and the parallel (physical) spaces, defined by:

\begin{equation}
    Q_{par}=\frac{1}{\sqrt{2(2+\tau)}} \begin{bmatrix} 0 & \tau & 1 & 0 & -\tau & 1 \\ \tau & 1 & 0 & -\tau & 1 & 0 \\ 1 & 0 & \tau & 1 & 0 & -\tau \end{bmatrix}
\end{equation}

\begin{equation}
    Q_{perp}=\frac{1}{\sqrt{2(2+\tau)}} \begin{bmatrix} 0 & -1 & \tau & 0 & 1 & \tau \\ -1 & \tau & 0 & 1 & \tau & 0 \\ \tau & 0 & -1 & \tau & 0 & 1 \end{bmatrix}
\end{equation}

\noindent where $\tau$ is the golden ratio. The icosahedral quasicrystal is generated by first projecting the 6D hypercubic body-centred lattice points onto the perpendicular plane. The cuts are done in this subspace, where we use a dodecahedron occupation domain to fix the same point group symmetries as in the icosahedral quasicrystal. The occupation domain is defined by the following vectors: $\pm 1/2(-1,1,1,1,1,1)$, $\pm 1/2(1,-1,1,1,1,1)$, $\pm 1/2(1,1,-1,1,1,1)$, $\pm 1/2(-1,1,1,-1,1,-1)$, $\pm 1/2(-1,1,-1,1,1,-1)$, $\pm 1/2(-1,-1,1,-1,1,1)$. The subset of lattice points inside the occupation domain are then projected from the 6D plane onto the parallel (physical) plane. The resulting points give the particle centers of the ideal quasicrystal.

To generate the $q/p$ approximant, one can use the same method but replace the $\tau$ by the ratio of two successive Fibonacci numbers $q/p$ in the matrix that defines the perpendicular space:

\begin{equation}
    Q_{perp}^{approx}=\frac{1}{\sqrt{2(p^2+q^2)}} \begin{bmatrix} 0 & -p & q & 0 & p & q \\ -p & q & 0 & p & q & 0 \\ q & 0 & -p & q & 0 & p \end{bmatrix}
\end{equation}

\subsection{Avoiding competing structures}

With the SAT-assembly framework one can test if different colouring designs are able to assemble or not competing structures. For a given total number of colours, one needs to extract the topology (similarly to the structures mentioned in the main text), then fix the patch colouring and colour interaction matrix variables to a given coloured design, and run the SAT-solver against the competing structure. If it is satisfied one can exclude this particular solution and find a new one for the original quasicrystal. This protocol can be repeated as many times as necessary to find a proper colouring. If the solution space is exhausted, one can change the total number of colours and restart. In Fig.~$S1$ we show a trajectory with $8$ species and $15$ colours, which was calculated without excluding any other competing structure, for the ideal quasicrystal that was simulated with $\rho=0.1$, $T=0.09$, and using $528$ particles. It is possible to observe that the inner shell of the quasicrystal is formed but then a subset of the species forms a stacking of hexagonal layers on top of it. Although we do not exclude this stacking specifically, after excluding the $BC8$ crystal we never observed the assembly of competing regular structures in simulations. The $BC8$ crystal topology that we explicitly exclude is included in Table~5.

\begin{figure}[t]
	\includegraphics{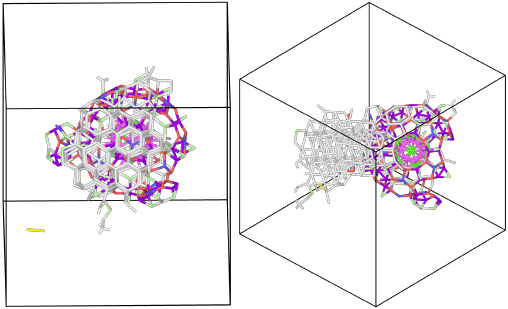}
	\caption{\label{figS1} Snapshots of a trajectory with $8$ species and $15$ colours for the ideal qusicrystal, with $\rho=0.1$, $T=0.09$, and using $528$ particles. Only the bonds between particles are shown through straight lines between particle centers. The left snapshot is along the symmetry axis of the hexagonal stacking while the right snapshot is along one of the five-fold axis of the quasicrystal cluster.}
\end{figure}

\begin{table*}
    \begin{center}
    \begin{tabular}{ cccc } 
    \hline
    Position $l_i$ & Slot $s_i$ & Position $l_j$ & Slot $s_j$ \\
    \hline
 0 & 1 & 9 & 1 \\
 0 & 2 & 12 & 1 \\
 0 & 3 & 10 & 1 \\
 0 & 0 & 15 & 0 \\
 1 & 1 & 3 & 1 \\
 1 & 2 & 5 & 1 \\
 1 & 3 & 7 & 1 \\
 1 & 0 & 2 & 0 \\
 2 & 1 & 4 & 1 \\
 2 & 2 & 8 & 1 \\
 2 & 3 & 6 & 1 \\
 3 & 2 & 12 & 3 \\
 3 & 3 & 9 & 2 \\
 3 & 0 & 4 & 0 \\
 4 & 2 & 14 & 1 \\
 4 & 3 & 11 & 1 \\
 5 & 2 & 10 & 3 \\
 5 & 3 & 12 & 2 \\
 5 & 0 & 6 & 0 \\
 6 & 2 & 11 & 2 \\
 6 & 3 & 13 & 1 \\
 7 & 2 & 9 & 3 \\
 7 & 3 & 10 & 2 \\
 7 & 0 & 8 & 0 \\
 8 & 2 & 13 & 2 \\
 8 & 3 & 14 & 3 \\
 9 & 0 & 14 & 0 \\
 10 & 0 & 13 & 0 \\
 11 & 3 & 15 & 1 \\
 11 & 0 & 12 & 0 \\
 13 & 3 & 15 & 3 \\
 14 & 2 & 15 & 2 \\
        \hline
        \end{tabular}
        \caption{$BC8$ crystal topology for SAT-assembly.}
    \end{center}
\end{table*}

\subsection{Two-step protocol quench for ideal quasicrystal assembly}

For the ideal quasicrystal we introduce a two-step quench protocol to aid the self-assembly process. We start at a higher temperature where only the patchy particles with $7$ and $6$ patches can remain bonded. This happens since these patches have the highest bonding probability and should bound to each other to form the $32$-particle icosahedral shell. Then, we instantly quench the system to the temperature of assembly where we observe that the quasicrystal is able to grow from this initial nucleus. Without this protocol, it is more challenging for the internal icosahedral shell to close, making the full assembly more difficult as well. Figure~$S2$ shows snapshots of the assembly process before and after the quench, to highlight how before the quench the icosahedral shell is stabilized and only the particles with many patches can remain bonded, but after the quench, all particles are able to form longer lasting bonds. 

\begin{figure}[t]
	\includegraphics{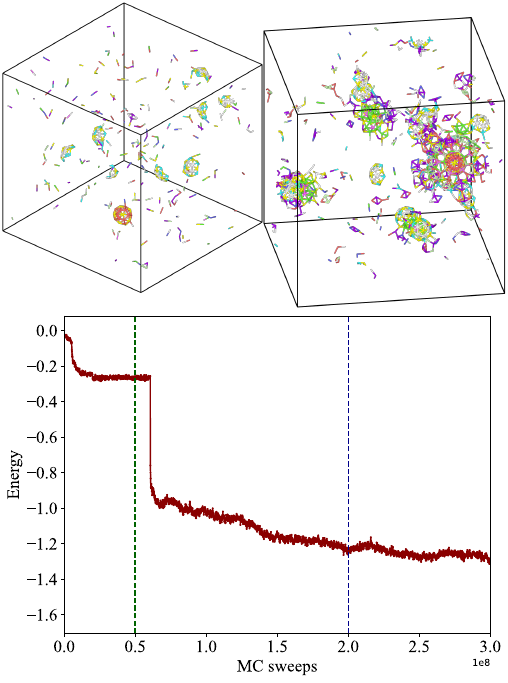}
	\caption{\label{figS2}Energy as a function of time for the ideal quasicrystal assembly with $28$ species and $27$ colours. Two configurations are highlighted, one before the quench with $T=0.12$ and another after the quench with $T=0.0875$. Before the quench only the particles with the most patches are able to remain bonded, which stabilizes more easily the $32$-particle icosahedral shell. After the quench, all particles are able to form longer lasting bonds and grow the quasicrystal from this initial nucleus.}
\end{figure}


\begin{thebibliography}{37}%
	\makeatletter
	\providecommand \@ifxundefined [1]{%
		\@ifx{#1\undefined}
	}%
	\providecommand \@ifnum [1]{%
		\ifnum #1\expandafter \@firstoftwo
		\else \expandafter \@secondoftwo
		\fi
	}%
	\providecommand \@ifx [1]{%
		\ifx #1\expandafter \@firstoftwo
		\else \expandafter \@secondoftwo
		\fi
	}%
	\providecommand \natexlab [1]{#1}%
	\providecommand \enquote  [1]{``#1''}%
	\providecommand \bibnamefont  [1]{#1}%
	\providecommand \bibfnamefont [1]{#1}%
	\providecommand \citenamefont [1]{#1}%
	\providecommand \href@noop [0]{\@secondoftwo}%
	\providecommand \href [0]{\begingroup \@sanitize@url \@href}%
	\providecommand \@href[1]{\@@startlink{#1}\@@href}%
	\providecommand \@@href[1]{\endgroup#1\@@endlink}%
	\providecommand \@sanitize@url [0]{\catcode `\\12\catcode `\$12\catcode
		`\&12\catcode `\#12\catcode `\^12\catcode `\_12\catcode `\%12\relax}%
	\providecommand \@@startlink[1]{}%
	\providecommand \@@endlink[0]{}%
	\providecommand \url  [0]{\begingroup\@sanitize@url \@url }%
	\providecommand \@url [1]{\endgroup\@href {#1}{\urlprefix }}%
	\providecommand \urlprefix  [0]{URL }%
	\providecommand \Eprint [0]{\href }%
	\providecommand \doibase [0]{https://doi.org/}%
	\providecommand \selectlanguage [0]{\@gobble}%
	\providecommand \bibinfo  [0]{\@secondoftwo}%
	\providecommand \bibfield  [0]{\@secondoftwo}%
	\providecommand \translation [1]{[#1]}%
	\providecommand \BibitemOpen [0]{}%
	\providecommand \bibitemStop [0]{}%
	\providecommand \bibitemNoStop [0]{.\EOS\space}%
	\providecommand \EOS [0]{\spacefactor3000\relax}%
	\providecommand \BibitemShut  [1]{\csname bibitem#1\endcsname}%
	\let\auto@bib@innerbib\@empty
	\bibitem [{\citenamefont {Posnjak}\ \emph {et~al.}(2024)\citenamefont
		{Posnjak}, \citenamefont {Yin}, \citenamefont {Butler}, \citenamefont
		{Bienek}, \citenamefont {Dass}, \citenamefont {Lee}, \citenamefont {Sharp},\
		and\ \citenamefont {Liedl}}]{Posnjak2024}%
	\BibitemOpen
	\bibfield  {author} {\bibinfo {author} {\bibfnamefont {G.}~\bibnamefont
			{Posnjak}}, \bibinfo {author} {\bibfnamefont {X.}~\bibnamefont {Yin}},
		\bibinfo {author} {\bibfnamefont {P.}~\bibnamefont {Butler}}, \bibinfo
		{author} {\bibfnamefont {O.}~\bibnamefont {Bienek}}, \bibinfo {author}
		{\bibfnamefont {M.}~\bibnamefont {Dass}}, \bibinfo {author} {\bibfnamefont
			{S.}~\bibnamefont {Lee}}, \bibinfo {author} {\bibfnamefont {I.~D.}\
			\bibnamefont {Sharp}},\ and\ \bibinfo {author} {\bibfnamefont
			{T.}~\bibnamefont {Liedl}},\ }\bibfield  {title} {\bibinfo {title}
		{Diamond-lattice photonic crystals assembled from dna origami},\ }\href
	{https://doi.org/10.1126/science.adl2733} {\bibfield  {journal} {\bibinfo
			{journal} {Science}\ }\textbf {\bibinfo {volume} {384}},\ \bibinfo {pages}
		{781} (\bibinfo {year} {2024})}\BibitemShut {NoStop}%
	\bibitem [{\citenamefont {Liu}\ \emph {et~al.}(2024)\citenamefont {Liu},
		\citenamefont {Matthies}, \citenamefont {Russo}, \citenamefont {Rovigatti},
		\citenamefont {Narayanan}, \citenamefont {Diep}, \citenamefont {McKeen},
		\citenamefont {Gang}, \citenamefont {Stephanopoulos}, \citenamefont
		{Sciortino}, \citenamefont {Yan}, \citenamefont {Romano},\ and\ \citenamefont
		{Šulc}}]{Liu2023}%
	\BibitemOpen
	\bibfield  {author} {\bibinfo {author} {\bibfnamefont {H.}~\bibnamefont
			{Liu}}, \bibinfo {author} {\bibfnamefont {M.}~\bibnamefont {Matthies}},
		\bibinfo {author} {\bibfnamefont {J.}~\bibnamefont {Russo}}, \bibinfo
		{author} {\bibfnamefont {L.}~\bibnamefont {Rovigatti}}, \bibinfo {author}
		{\bibfnamefont {R.~P.}\ \bibnamefont {Narayanan}}, \bibinfo {author}
		{\bibfnamefont {T.}~\bibnamefont {Diep}}, \bibinfo {author} {\bibfnamefont
			{D.}~\bibnamefont {McKeen}}, \bibinfo {author} {\bibfnamefont
			{O.}~\bibnamefont {Gang}}, \bibinfo {author} {\bibfnamefont {N.}~\bibnamefont
			{Stephanopoulos}}, \bibinfo {author} {\bibfnamefont {F.}~\bibnamefont
			{Sciortino}}, \bibinfo {author} {\bibfnamefont {H.}~\bibnamefont {Yan}},
		\bibinfo {author} {\bibfnamefont {F.}~\bibnamefont {Romano}},\ and\ \bibinfo
		{author} {\bibfnamefont {P.}~\bibnamefont {Šulc}},\ }\bibfield  {title}
	{\bibinfo {title} {Inverse design of a pyrochlore lattice of dna origami
			through model-driven experiments},\ }\href
	{https://doi.org/10.1126/science.adl5549} {\bibfield  {journal} {\bibinfo
			{journal} {Science}\ }\textbf {\bibinfo {volume} {384}},\ \bibinfo {pages}
		{776} (\bibinfo {year} {2024})}\BibitemShut {NoStop}%
	\bibitem [{\citenamefont {Macfarlane}\ \emph {et~al.}(2011)\citenamefont
		{Macfarlane}, \citenamefont {Lee}, \citenamefont {Jones}, \citenamefont
		{Harris}, \citenamefont {Schatz},\ and\ \citenamefont
		{Mirkin}}]{Macfarlane2011}%
	\BibitemOpen
	\bibfield  {author} {\bibinfo {author} {\bibfnamefont {R.~J.}\ \bibnamefont
			{Macfarlane}}, \bibinfo {author} {\bibfnamefont {B.}~\bibnamefont {Lee}},
		\bibinfo {author} {\bibfnamefont {M.~R.}\ \bibnamefont {Jones}}, \bibinfo
		{author} {\bibfnamefont {N.}~\bibnamefont {Harris}}, \bibinfo {author}
		{\bibfnamefont {G.~C.}\ \bibnamefont {Schatz}},\ and\ \bibinfo {author}
		{\bibfnamefont {C.~A.}\ \bibnamefont {Mirkin}},\ }\bibfield  {title}
	{\bibinfo {title} {Nanoparticle superlattice engineering with dna},\ }\href
	{https://doi.org/10.1126/science.1210493} {\bibfield  {journal} {\bibinfo
			{journal} {Science}\ }\textbf {\bibinfo {volume} {334}},\ \bibinfo {pages}
		{204} (\bibinfo {year} {2011})}\BibitemShut {NoStop}%
	\bibitem [{\citenamefont {Jones}\ \emph {et~al.}(2015)\citenamefont {Jones},
		\citenamefont {Seeman},\ and\ \citenamefont {Mirkin}}]{Jones2015}%
	\BibitemOpen
	\bibfield  {author} {\bibinfo {author} {\bibfnamefont {M.~R.}\ \bibnamefont
			{Jones}}, \bibinfo {author} {\bibfnamefont {N.~C.}\ \bibnamefont {Seeman}},\
		and\ \bibinfo {author} {\bibfnamefont {C.~A.}\ \bibnamefont {Mirkin}},\
	}\bibfield  {title} {\bibinfo {title} {Programmable materials and the nature
			of the dna bond},\ }\href {https://doi.org/10.1126/science.1260901}
	{\bibfield  {journal} {\bibinfo  {journal} {Science}\ }\textbf {\bibinfo
			{volume} {347}},\ \bibinfo {pages} {1260901} (\bibinfo {year}
		{2015})}\BibitemShut {NoStop}%
	\bibitem [{\citenamefont {Romano}\ \emph {et~al.}(2011)\citenamefont {Romano},
		\citenamefont {Sanz},\ and\ \citenamefont {Sciortino}}]{Romano2011}%
	\BibitemOpen
	\bibfield  {author} {\bibinfo {author} {\bibfnamefont {F.}~\bibnamefont
			{Romano}}, \bibinfo {author} {\bibfnamefont {E.}~\bibnamefont {Sanz}},\ and\
		\bibinfo {author} {\bibfnamefont {F.}~\bibnamefont {Sciortino}},\ }\bibfield
	{title} {\bibinfo {title} {{Crystallization of tetrahedral patchy particles
				in silico}},\ }\href {https://doi.org/10.1063/1.3578182} {\bibfield
		{journal} {\bibinfo  {journal} {The Journal of Chemical Physics}\ }\textbf
		{\bibinfo {volume} {134}},\ \bibinfo {pages} {174502} (\bibinfo {year}
		{2011})}\BibitemShut {NoStop}%
	\bibitem [{\citenamefont {Wang}\ \emph {et~al.}(2012)\citenamefont {Wang},
		\citenamefont {Wang}, \citenamefont {Breed}, \citenamefont {Manoharan},
		\citenamefont {Feng}, \citenamefont {Hollingsworth}, \citenamefont {Weck},\
		and\ \citenamefont {Pine}}]{Wang2012}%
	\BibitemOpen
	\bibfield  {author} {\bibinfo {author} {\bibfnamefont {Y.}~\bibnamefont
			{Wang}}, \bibinfo {author} {\bibfnamefont {Y.}~\bibnamefont {Wang}}, \bibinfo
		{author} {\bibfnamefont {D.~R.}\ \bibnamefont {Breed}}, \bibinfo {author}
		{\bibfnamefont {V.~N.}\ \bibnamefont {Manoharan}}, \bibinfo {author}
		{\bibfnamefont {L.}~\bibnamefont {Feng}}, \bibinfo {author} {\bibfnamefont
			{A.~D.}\ \bibnamefont {Hollingsworth}}, \bibinfo {author} {\bibfnamefont
			{M.}~\bibnamefont {Weck}},\ and\ \bibinfo {author} {\bibfnamefont {D.~J.}\
			\bibnamefont {Pine}},\ }\bibfield  {title} {\bibinfo {title} {Colloids with
			valence and specific directional bonding},\ }\href@noop {} {\bibfield
		{journal} {\bibinfo  {journal} {Nature}\ }\textbf {\bibinfo {volume} {491}},\
		\bibinfo {pages} {51} (\bibinfo {year} {2012})}\BibitemShut {NoStop}%
	\bibitem [{\citenamefont {Videbæk}\ \emph {et~al.}(2024)\citenamefont
		{Videbæk}, \citenamefont {Hayakawa}, \citenamefont {Grason}, \citenamefont
		{Hagan}, \citenamefont {Fraden},\ and\ \citenamefont
		{Rogers}}]{Videbaek2024}%
	\BibitemOpen
	\bibfield  {author} {\bibinfo {author} {\bibfnamefont {T.~E.}\ \bibnamefont
			{Videbæk}}, \bibinfo {author} {\bibfnamefont {D.}~\bibnamefont {Hayakawa}},
		\bibinfo {author} {\bibfnamefont {G.~M.}\ \bibnamefont {Grason}}, \bibinfo
		{author} {\bibfnamefont {M.~F.}\ \bibnamefont {Hagan}}, \bibinfo {author}
		{\bibfnamefont {S.}~\bibnamefont {Fraden}},\ and\ \bibinfo {author}
		{\bibfnamefont {W.~B.}\ \bibnamefont {Rogers}},\ }\bibfield  {title}
	{\bibinfo {title} {Economical routes to size-specific assembly of
			self-closing structures},\ }\href {https://doi.org/10.1126/sciadv.ado5979}
	{\bibfield  {journal} {\bibinfo  {journal} {Science Advances}\ }\textbf
		{\bibinfo {volume} {10}},\ \bibinfo {pages} {eado5979} (\bibinfo {year}
		{2024})}\BibitemShut {NoStop}%
	\bibitem [{\citenamefont {Flavell}\ \emph {et~al.}(2023)\citenamefont
		{Flavell}, \citenamefont {Neophytou}, \citenamefont {Demetriadou},
		\citenamefont {Albrecht},\ and\ \citenamefont {Chakrabarti}}]{Flavell2023}%
	\BibitemOpen
	\bibfield  {author} {\bibinfo {author} {\bibfnamefont {W.}~\bibnamefont
			{Flavell}}, \bibinfo {author} {\bibfnamefont {A.}~\bibnamefont {Neophytou}},
		\bibinfo {author} {\bibfnamefont {A.}~\bibnamefont {Demetriadou}}, \bibinfo
		{author} {\bibfnamefont {T.}~\bibnamefont {Albrecht}},\ and\ \bibinfo
		{author} {\bibfnamefont {D.}~\bibnamefont {Chakrabarti}},\ }\bibfield
	{title} {\bibinfo {title} {Programmed self-assembly of single colloidal
			gyroids for chiral photonic crystals},\ }\href
	{https://doi.org/https://doi.org/10.1002/adma.202211197} {\bibfield
		{journal} {\bibinfo  {journal} {Advanced Materials}\ }\textbf {\bibinfo
			{volume} {35}},\ \bibinfo {pages} {2211197} (\bibinfo {year}
		{2023})}\BibitemShut {NoStop}%
	\bibitem [{\citenamefont {King}\ \emph {et~al.}(2024)\citenamefont {King},
		\citenamefont {Du}, \citenamefont {Zhu}, \citenamefont {Schoenholz},\ and\
		\citenamefont {Brenner}}]{King2024}%
	\BibitemOpen
	\bibfield  {author} {\bibinfo {author} {\bibfnamefont {E.~M.}\ \bibnamefont
			{King}}, \bibinfo {author} {\bibfnamefont {C.~X.}\ \bibnamefont {Du}},
		\bibinfo {author} {\bibfnamefont {Q.-Z.}\ \bibnamefont {Zhu}}, \bibinfo
		{author} {\bibfnamefont {S.~S.}\ \bibnamefont {Schoenholz}},\ and\ \bibinfo
		{author} {\bibfnamefont {M.~P.}\ \bibnamefont {Brenner}},\ }\bibfield
	{title} {\bibinfo {title} {Programming patchy particles for materials
			assembly design},\ }\href {https://doi.org/10.1073/pnas.2311891121}
	{\bibfield  {journal} {\bibinfo  {journal} {Proceedings of the National
				Academy of Sciences}\ }\textbf {\bibinfo {volume} {121}},\ \bibinfo {pages}
		{e2311891121} (\bibinfo {year} {2024})}\BibitemShut {NoStop}%
	\bibitem [{\citenamefont {Coli}\ \emph {et~al.}(2022)\citenamefont {Coli},
		\citenamefont {Boattini}, \citenamefont {Filion},\ and\ \citenamefont
		{Dijkstra}}]{Coli2022}%
	\BibitemOpen
	\bibfield  {author} {\bibinfo {author} {\bibfnamefont {G.~M.}\ \bibnamefont
			{Coli}}, \bibinfo {author} {\bibfnamefont {E.}~\bibnamefont {Boattini}},
		\bibinfo {author} {\bibfnamefont {L.}~\bibnamefont {Filion}},\ and\ \bibinfo
		{author} {\bibfnamefont {M.}~\bibnamefont {Dijkstra}},\ }\bibfield  {title}
	{\bibinfo {title} {Inverse design of soft materials via a deep
			learning–based evolutionary strategy},\ }\href
	{https://doi.org/10.1126/sciadv.abj6731} {\bibfield  {journal} {\bibinfo
			{journal} {Science Advances}\ }\textbf {\bibinfo {volume} {8}},\ \bibinfo
		{pages} {eabj6731} (\bibinfo {year} {2022})}\BibitemShut {NoStop}%
	\bibitem [{\citenamefont {Goldman}\ and\ \citenamefont
		{Kelton}(1993)}]{Goldman1993}%
	\BibitemOpen
	\bibfield  {author} {\bibinfo {author} {\bibfnamefont {A.~I.}\ \bibnamefont
			{Goldman}}\ and\ \bibinfo {author} {\bibfnamefont {R.~F.}\ \bibnamefont
			{Kelton}},\ }\bibfield  {title} {\bibinfo {title} {Quasicrystals and
			crystalline approximants},\ }\href
	{https://doi.org/10.1103/RevModPhys.65.213} {\bibfield  {journal} {\bibinfo
			{journal} {Rev. Mod. Phys.}\ }\textbf {\bibinfo {volume} {65}},\ \bibinfo
		{pages} {213} (\bibinfo {year} {1993})}\BibitemShut {NoStop}%
	\bibitem [{\citenamefont {Engel}\ \emph {et~al.}(2015)\citenamefont {Engel},
		\citenamefont {Damasceno}, \citenamefont {Phillips},\ and\ \citenamefont
		{Glotzer}}]{Engel2015}%
	\BibitemOpen
	\bibfield  {author} {\bibinfo {author} {\bibfnamefont {M.}~\bibnamefont
			{Engel}}, \bibinfo {author} {\bibfnamefont {P.~F.}\ \bibnamefont
			{Damasceno}}, \bibinfo {author} {\bibfnamefont {C.~L.}\ \bibnamefont
			{Phillips}},\ and\ \bibinfo {author} {\bibfnamefont {S.~C.}\ \bibnamefont
			{Glotzer}},\ }\bibfield  {title} {\bibinfo {title} {Computational
			self-assembly of a one-component icosahedral quasicrystal},\ }\href@noop {}
	{\bibfield  {journal} {\bibinfo  {journal} {Nature materials}\ }\textbf
		{\bibinfo {volume} {14}},\ \bibinfo {pages} {109} (\bibinfo {year}
		{2015})}\BibitemShut {NoStop}%
	\bibitem [{\citenamefont {Noya}\ \emph {et~al.}(2021)\citenamefont {Noya},
		\citenamefont {Wong}, \citenamefont {Llombart},\ and\ \citenamefont
		{Doye}}]{Noya2021}%
	\BibitemOpen
	\bibfield  {author} {\bibinfo {author} {\bibfnamefont {E.~G.}\ \bibnamefont
			{Noya}}, \bibinfo {author} {\bibfnamefont {C.~K.}\ \bibnamefont {Wong}},
		\bibinfo {author} {\bibfnamefont {P.}~\bibnamefont {Llombart}},\ and\
		\bibinfo {author} {\bibfnamefont {J.~P.}\ \bibnamefont {Doye}},\ }\bibfield
	{title} {\bibinfo {title} {{How to design an icosahedral quasicrystal through
				directional bonding}},\ }\href {https://doi.org/10.1038/s41586-021-03700-2}
	{\bibfield  {journal} {\bibinfo  {journal} {Nature}\ }\textbf {\bibinfo
			{volume} {596}},\ \bibinfo {pages} {367} (\bibinfo {year}
		{2021})}\BibitemShut {NoStop}%
	\bibitem [{\citenamefont {Zhou}\ \emph {et~al.}(2024)\citenamefont {Zhou},
		\citenamefont {Lim}, \citenamefont {Lin}, \citenamefont {Lee}, \citenamefont
		{Li}, \citenamefont {Huang}, \citenamefont {Du}, \citenamefont {Lee},
		\citenamefont {Wang}, \citenamefont {S{\'a}nchez-Iglesias} \emph
		{et~al.}}]{Zhou2024}%
	\BibitemOpen
	\bibfield  {author} {\bibinfo {author} {\bibfnamefont {W.}~\bibnamefont
			{Zhou}}, \bibinfo {author} {\bibfnamefont {Y.}~\bibnamefont {Lim}}, \bibinfo
		{author} {\bibfnamefont {H.}~\bibnamefont {Lin}}, \bibinfo {author}
		{\bibfnamefont {S.}~\bibnamefont {Lee}}, \bibinfo {author} {\bibfnamefont
			{Y.}~\bibnamefont {Li}}, \bibinfo {author} {\bibfnamefont {Z.}~\bibnamefont
			{Huang}}, \bibinfo {author} {\bibfnamefont {J.~S.}\ \bibnamefont {Du}},
		\bibinfo {author} {\bibfnamefont {B.}~\bibnamefont {Lee}}, \bibinfo {author}
		{\bibfnamefont {S.}~\bibnamefont {Wang}}, \bibinfo {author} {\bibfnamefont
			{A.}~\bibnamefont {S{\'a}nchez-Iglesias}}, \emph {et~al.},\ }\bibfield
	{title} {\bibinfo {title} {Colloidal quasicrystals engineered with dna},\
	}\href@noop {} {\bibfield  {journal} {\bibinfo  {journal} {Nature materials}\
		}\textbf {\bibinfo {volume} {23}},\ \bibinfo {pages} {424} (\bibinfo {year}
		{2024})}\BibitemShut {NoStop}%
	\bibitem [{\citenamefont {Plati}\ \emph {et~al.}(2024)\citenamefont {Plati},
		\citenamefont {Maire}, \citenamefont {Fayen}, \citenamefont {Boulogne},
		\citenamefont {Restagno}, \citenamefont {Smallenburg},\ and\ \citenamefont
		{Foffi}}]{Plati2024}%
	\BibitemOpen
	\bibfield  {author} {\bibinfo {author} {\bibfnamefont {A.}~\bibnamefont
			{Plati}}, \bibinfo {author} {\bibfnamefont {R.}~\bibnamefont {Maire}},
		\bibinfo {author} {\bibfnamefont {E.}~\bibnamefont {Fayen}}, \bibinfo
		{author} {\bibfnamefont {F.}~\bibnamefont {Boulogne}}, \bibinfo {author}
		{\bibfnamefont {F.}~\bibnamefont {Restagno}}, \bibinfo {author}
		{\bibfnamefont {F.}~\bibnamefont {Smallenburg}},\ and\ \bibinfo {author}
		{\bibfnamefont {G.}~\bibnamefont {Foffi}},\ }\bibfield  {title} {\bibinfo
		{title} {Quasi-crystalline order in vibrating granular matter},\ }\href@noop
	{} {\bibfield  {journal} {\bibinfo  {journal} {Nature Physics}\ }\textbf
		{\bibinfo {volume} {20}},\ \bibinfo {pages} {1} (\bibinfo {year}
		{2024})}\BibitemShut {NoStop}%
	\bibitem [{\citenamefont {Lee}\ and\ \citenamefont {Glotzer}(2022)}]{Lee2022}%
	\BibitemOpen
	\bibfield  {author} {\bibinfo {author} {\bibfnamefont {S.}~\bibnamefont
			{Lee}}\ and\ \bibinfo {author} {\bibfnamefont {S.~C.}\ \bibnamefont
			{Glotzer}},\ }\bibfield  {title} {\bibinfo {title} {Entropically engineered
			formation of fivefold and icosahedral twinned clusters of colloidal shapes},\
	}\href@noop {} {\bibfield  {journal} {\bibinfo  {journal} {Nature
				Communications}\ }\textbf {\bibinfo {volume} {13}},\ \bibinfo {pages} {7362}
		(\bibinfo {year} {2022})}\BibitemShut {NoStop}%
	\bibitem [{\citenamefont {Fayen}\ \emph {et~al.}(2024)\citenamefont {Fayen},
		\citenamefont {Filion}, \citenamefont {Foffi},\ and\ \citenamefont
		{Smallenburg}}]{Fayen2024}%
	\BibitemOpen
	\bibfield  {author} {\bibinfo {author} {\bibfnamefont {E.}~\bibnamefont
			{Fayen}}, \bibinfo {author} {\bibfnamefont {L.}~\bibnamefont {Filion}},
		\bibinfo {author} {\bibfnamefont {G.}~\bibnamefont {Foffi}},\ and\ \bibinfo
		{author} {\bibfnamefont {F.}~\bibnamefont {Smallenburg}},\ }\bibfield
	{title} {\bibinfo {title} {Quasicrystal of binary hard spheres on a plane
			stabilized by configurational entropy},\ }\href
	{https://doi.org/10.1103/PhysRevLett.132.048202} {\bibfield  {journal}
		{\bibinfo  {journal} {Phys. Rev. Lett.}\ }\textbf {\bibinfo {volume} {132}},\
		\bibinfo {pages} {048202} (\bibinfo {year} {2024})}\BibitemShut {NoStop}%
	\bibitem [{\citenamefont {Russo}\ \emph {et~al.}(2022)\citenamefont {Russo},
		\citenamefont {Romano}, \citenamefont {Kroc}, \citenamefont {Sciortino},
		\citenamefont {Rovigatti},\ and\ \citenamefont {{\v{S}}ulc}}]{Russo2022}%
	\BibitemOpen
	\bibfield  {author} {\bibinfo {author} {\bibfnamefont {J.}~\bibnamefont
			{Russo}}, \bibinfo {author} {\bibfnamefont {F.}~\bibnamefont {Romano}},
		\bibinfo {author} {\bibfnamefont {L.}~\bibnamefont {Kroc}}, \bibinfo {author}
		{\bibfnamefont {F.}~\bibnamefont {Sciortino}}, \bibinfo {author}
		{\bibfnamefont {L.}~\bibnamefont {Rovigatti}},\ and\ \bibinfo {author}
		{\bibfnamefont {P.}~\bibnamefont {{\v{S}}ulc}},\ }\bibfield  {title}
	{\bibinfo {title} {{SAT}-assembly: a new approach for designing
			self-assembling systems},\ }\href {https://doi.org/10.1088/1361-648x/ac5479}
	{\bibfield  {journal} {\bibinfo  {journal} {Journal of Physics: Condensed
				Matter}\ }\textbf {\bibinfo {volume} {34}},\ \bibinfo {pages} {354002}
		(\bibinfo {year} {2022})}\BibitemShut {NoStop}%
	\bibitem [{\citenamefont {Sigl}\ \emph {et~al.}(2021)\citenamefont {Sigl},
		\citenamefont {Willner}, \citenamefont {Engelen}, \citenamefont {Kretzmann},
		\citenamefont {Sachenbacher}, \citenamefont {Liedl}, \citenamefont {Kolbe},
		\citenamefont {Wilsch}, \citenamefont {Aghvami}, \citenamefont {Protzer}
		\emph {et~al.}}]{sigl2021programmable}%
	\BibitemOpen
	\bibfield  {author} {\bibinfo {author} {\bibfnamefont {C.}~\bibnamefont
			{Sigl}}, \bibinfo {author} {\bibfnamefont {E.~M.}\ \bibnamefont {Willner}},
		\bibinfo {author} {\bibfnamefont {W.}~\bibnamefont {Engelen}}, \bibinfo
		{author} {\bibfnamefont {J.~A.}\ \bibnamefont {Kretzmann}}, \bibinfo {author}
		{\bibfnamefont {K.}~\bibnamefont {Sachenbacher}}, \bibinfo {author}
		{\bibfnamefont {A.}~\bibnamefont {Liedl}}, \bibinfo {author} {\bibfnamefont
			{F.}~\bibnamefont {Kolbe}}, \bibinfo {author} {\bibfnamefont
			{F.}~\bibnamefont {Wilsch}}, \bibinfo {author} {\bibfnamefont {S.~A.}\
			\bibnamefont {Aghvami}}, \bibinfo {author} {\bibfnamefont {U.}~\bibnamefont
			{Protzer}}, \emph {et~al.},\ }\bibfield  {title} {\bibinfo {title}
		{Programmable icosahedral shell system for virus trapping},\ }\href@noop {}
	{\bibfield  {journal} {\bibinfo  {journal} {Nature materials}\ }\textbf
		{\bibinfo {volume} {20}},\ \bibinfo {pages} {1281} (\bibinfo {year}
		{2021})}\BibitemShut {NoStop}%
	\bibitem [{\citenamefont {Mosayebi}\ \emph {et~al.}(2017)\citenamefont
		{Mosayebi}, \citenamefont {Shoemark}, \citenamefont {Fletcher}, \citenamefont
		{Sessions}, \citenamefont {Linden}, \citenamefont {Woolfson},\ and\
		\citenamefont {Liverpool}}]{Mosayebi2017}%
	\BibitemOpen
	\bibfield  {author} {\bibinfo {author} {\bibfnamefont {M.}~\bibnamefont
			{Mosayebi}}, \bibinfo {author} {\bibfnamefont {D.~K.}\ \bibnamefont
			{Shoemark}}, \bibinfo {author} {\bibfnamefont {J.~M.}\ \bibnamefont
			{Fletcher}}, \bibinfo {author} {\bibfnamefont {R.~B.}\ \bibnamefont
			{Sessions}}, \bibinfo {author} {\bibfnamefont {N.}~\bibnamefont {Linden}},
		\bibinfo {author} {\bibfnamefont {D.~N.}\ \bibnamefont {Woolfson}},\ and\
		\bibinfo {author} {\bibfnamefont {T.~B.}\ \bibnamefont {Liverpool}},\
	}\bibfield  {title} {\bibinfo {title} {{Beyond icosahedral symmetry in
				packings of proteins in spherical shells}},\ }\href
	{https://doi.org/10.1073/PNAS.1706825114/SUPPL_FILE/PNAS.1706825114.SAPP.PDF}
	{\bibfield  {journal} {\bibinfo  {journal} {Proceedings of the National
				Academy of Sciences of the United States of America}\ }\textbf {\bibinfo
			{volume} {114}},\ \bibinfo {pages} {9014} (\bibinfo {year}
		{2017})}\BibitemShut {NoStop}%
	\bibitem [{\citenamefont {Lee}\ \emph {et~al.}(2022)\citenamefont {Lee},
		\citenamefont {Kim}, \citenamefont {Lee},\ and\ \citenamefont
		{Kim}}]{Lee2022a}%
	\BibitemOpen
	\bibfield  {author} {\bibinfo {author} {\bibfnamefont {J.~G.}\ \bibnamefont
			{Lee}}, \bibinfo {author} {\bibfnamefont {K.~S.}\ \bibnamefont {Kim}},
		\bibinfo {author} {\bibfnamefont {J.~Y.}\ \bibnamefont {Lee}},\ and\ \bibinfo
		{author} {\bibfnamefont {D.-N.}\ \bibnamefont {Kim}},\ }\bibfield  {title}
	{\bibinfo {title} {{Predicting the Free-Form Shape of Structured DNA
				Assemblies from Their Lattice-Based Design Blueprint}},\ }\href
	{https://doi.org/10.1021/acsnano.1c10347} {\bibfield  {journal} {\bibinfo
			{journal} {ACS Nano}\ }\textbf {\bibinfo {volume} {16}},\ \bibinfo {pages}
		{4289} (\bibinfo {year} {2022})}\BibitemShut {NoStop}%
	\bibitem [{\citenamefont {Jun}\ \emph {et~al.}(2021)\citenamefont {Jun},
		\citenamefont {Wang}, \citenamefont {Parsons}, \citenamefont {Bricker},
		\citenamefont {John}, \citenamefont {Li}, \citenamefont {Jackson},
		\citenamefont {Chiu},\ and\ \citenamefont {Bathe}}]{Jun2021}%
	\BibitemOpen
	\bibfield  {author} {\bibinfo {author} {\bibfnamefont {H.}~\bibnamefont
			{Jun}}, \bibinfo {author} {\bibfnamefont {X.}~\bibnamefont {Wang}}, \bibinfo
		{author} {\bibfnamefont {M.~F.}\ \bibnamefont {Parsons}}, \bibinfo {author}
		{\bibfnamefont {W.~P.}\ \bibnamefont {Bricker}}, \bibinfo {author}
		{\bibfnamefont {T.}~\bibnamefont {John}}, \bibinfo {author} {\bibfnamefont
			{S.}~\bibnamefont {Li}}, \bibinfo {author} {\bibfnamefont {S.}~\bibnamefont
			{Jackson}}, \bibinfo {author} {\bibfnamefont {W.}~\bibnamefont {Chiu}},\ and\
		\bibinfo {author} {\bibfnamefont {M.}~\bibnamefont {Bathe}},\ }\bibfield
	{title} {\bibinfo {title} {{Rapid prototyping of arbitrary 2D and 3D
				wireframe DNA origami}},\ }\href {https://doi.org/10.1093/nar/gkab762}
	{\bibfield  {journal} {\bibinfo  {journal} {Nucleic Acids Research}\ }\textbf
		{\bibinfo {volume} {49}},\ \bibinfo {pages} {10265} (\bibinfo {year}
		{2021})}\BibitemShut {NoStop}%
	\bibitem [{\citenamefont {Rothemund}(2006)}]{Rothemund2006}%
	\BibitemOpen
	\bibfield  {author} {\bibinfo {author} {\bibfnamefont {P.~W.}\ \bibnamefont
			{Rothemund}},\ }\bibfield  {title} {\bibinfo {title} {{Folding DNA to create
				nanoscale shapes and patterns}},\ }\href
	{https://doi.org/10.1038/nature04586} {\bibfield  {journal} {\bibinfo
			{journal} {Nature}\ }\textbf {\bibinfo {volume} {440}},\ \bibinfo {pages}
		{297} (\bibinfo {year} {2006})}\BibitemShut {NoStop}%
	\bibitem [{\citenamefont {Sacanna}\ and\ \citenamefont
		{Pine}(2011)}]{Sacanna2011}%
	\BibitemOpen
	\bibfield  {author} {\bibinfo {author} {\bibfnamefont {S.}~\bibnamefont
			{Sacanna}}\ and\ \bibinfo {author} {\bibfnamefont {D.~J.}\ \bibnamefont
			{Pine}},\ }\bibfield  {title} {\bibinfo {title} {Shape-anisotropic colloids:
			Building blocks for complex assemblies},\ }\href
	{https://doi.org/10.1016/j.cocis.2011.01.003} {\bibfield  {journal} {\bibinfo
			{journal} {Current Opinion in Colloid \& Interface Science}\ }\textbf
		{\bibinfo {volume} {16}},\ \bibinfo {pages} {96} (\bibinfo {year}
		{2011})}\BibitemShut {NoStop}%
	\bibitem [{\citenamefont {Wang}\ \emph {et~al.}(2015)\citenamefont {Wang},
		\citenamefont {Wang}, \citenamefont {Zheng}, \citenamefont {Ducrot},
		\citenamefont {Yodh}, \citenamefont {Weck},\ and\ \citenamefont
		{Pine}}]{Wang2015}%
	\BibitemOpen
	\bibfield  {author} {\bibinfo {author} {\bibfnamefont {Y.}~\bibnamefont
			{Wang}}, \bibinfo {author} {\bibfnamefont {Y.}~\bibnamefont {Wang}}, \bibinfo
		{author} {\bibfnamefont {X.}~\bibnamefont {Zheng}}, \bibinfo {author}
		{\bibfnamefont {{\'{E}}.}~\bibnamefont {Ducrot}}, \bibinfo {author}
		{\bibfnamefont {J.~S.}\ \bibnamefont {Yodh}}, \bibinfo {author}
		{\bibfnamefont {M.}~\bibnamefont {Weck}},\ and\ \bibinfo {author}
		{\bibfnamefont {D.~J.}\ \bibnamefont {Pine}},\ }\bibfield  {title} {\bibinfo
		{title} {{Crystallization of DNA-coated colloids}},\ }\href
	{https://doi.org/10.1038/ncomms8253} {\bibfield  {journal} {\bibinfo
			{journal} {Nature Communications}\ }\textbf {\bibinfo {volume} {6}},\
		\bibinfo {pages} {7253} (\bibinfo {year} {2015})}\BibitemShut {NoStop}%
	\bibitem [{\citenamefont {Rovigatti}\ \emph {et~al.}(2022)\citenamefont
		{Rovigatti}, \citenamefont {Russo}, \citenamefont {Romano}, \citenamefont
		{Matthies}, \citenamefont {Kroc},\ and\ \citenamefont
		{Šulc}}]{Rovigatti2022}%
	\BibitemOpen
	\bibfield  {author} {\bibinfo {author} {\bibfnamefont {L.}~\bibnamefont
			{Rovigatti}}, \bibinfo {author} {\bibfnamefont {J.}~\bibnamefont {Russo}},
		\bibinfo {author} {\bibfnamefont {F.}~\bibnamefont {Romano}}, \bibinfo
		{author} {\bibfnamefont {M.}~\bibnamefont {Matthies}}, \bibinfo {author}
		{\bibfnamefont {L.}~\bibnamefont {Kroc}},\ and\ \bibinfo {author}
		{\bibfnamefont {P.}~\bibnamefont {Šulc}},\ }\bibfield  {title} {\bibinfo
		{title} {A simple solution to the problem of self-assembling cubic diamond
			crystals},\ }\href {https://doi.org/10.1039/D2NR03533B} {\bibfield  {journal}
		{\bibinfo  {journal} {Nanoscale}\ }\textbf {\bibinfo {volume} {14}},\
		\bibinfo {pages} {14268} (\bibinfo {year} {2022})}\BibitemShut {NoStop}%
	\bibitem [{\citenamefont {Ravaine}\ and\ \citenamefont
		{Duguet}(2017)}]{ravaine2017synthesis}%
	\BibitemOpen
	\bibfield  {author} {\bibinfo {author} {\bibfnamefont {S.}~\bibnamefont
			{Ravaine}}\ and\ \bibinfo {author} {\bibfnamefont {E.}~\bibnamefont
			{Duguet}},\ }\bibfield  {title} {\bibinfo {title} {Synthesis and assembly of
			patchy particles: Recent progress and future prospects},\ }\href@noop {}
	{\bibfield  {journal} {\bibinfo  {journal} {Current Opinion in Colloid \&
				Interface Science}\ }\textbf {\bibinfo {volume} {30}},\ \bibinfo {pages} {45}
		(\bibinfo {year} {2017})}\BibitemShut {NoStop}%
	\bibitem [{\citenamefont {Pawar}\ and\ \citenamefont
		{Kretzschmar}(2010)}]{pawar2010fabrication}%
	\BibitemOpen
	\bibfield  {author} {\bibinfo {author} {\bibfnamefont {A.~B.}\ \bibnamefont
			{Pawar}}\ and\ \bibinfo {author} {\bibfnamefont {I.}~\bibnamefont
			{Kretzschmar}},\ }\bibfield  {title} {\bibinfo {title} {Fabrication,
			assembly, and application of patchy particles},\ }\href@noop {} {\bibfield
		{journal} {\bibinfo  {journal} {Macromolecular rapid communications}\
		}\textbf {\bibinfo {volume} {31}},\ \bibinfo {pages} {150} (\bibinfo {year}
		{2010})}\BibitemShut {NoStop}%
	\bibitem [{\citenamefont {Zhang}\ \emph {et~al.}(2022)\citenamefont {Zhang},
		\citenamefont {Xu}, \citenamefont {Huang}, \citenamefont {Sun}, \citenamefont
		{Liu}, \citenamefont {Wan}, \citenamefont {Chen}, \citenamefont {Yang},
		\citenamefont {Yang},\ and\ \citenamefont {Song}}]{Zhang2022}%
	\BibitemOpen
	\bibfield  {author} {\bibinfo {author} {\bibfnamefont {J.}~\bibnamefont
			{Zhang}}, \bibinfo {author} {\bibfnamefont {Y.}~\bibnamefont {Xu}}, \bibinfo
		{author} {\bibfnamefont {Y.}~\bibnamefont {Huang}}, \bibinfo {author}
		{\bibfnamefont {M.}~\bibnamefont {Sun}}, \bibinfo {author} {\bibfnamefont
			{S.}~\bibnamefont {Liu}}, \bibinfo {author} {\bibfnamefont {S.}~\bibnamefont
			{Wan}}, \bibinfo {author} {\bibfnamefont {H.}~\bibnamefont {Chen}}, \bibinfo
		{author} {\bibfnamefont {C.}~\bibnamefont {Yang}}, \bibinfo {author}
		{\bibfnamefont {Y.}~\bibnamefont {Yang}},\ and\ \bibinfo {author}
		{\bibfnamefont {Y.}~\bibnamefont {Song}},\ }\bibfield  {title} {\bibinfo
		{title} {Spatially patterned neutralizing icosahedral dna nanocage for
			efficient sars-cov-2 blocking},\ }\href
	{https://doi.org/10.1021/jacs.2c02764} {\bibfield  {journal} {\bibinfo
			{journal} {Journal of the American Chemical Society}\ }\textbf {\bibinfo
			{volume} {144}},\ \bibinfo {pages} {13146} (\bibinfo {year}
		{2022})}\BibitemShut {NoStop}%
	\bibitem [{\citenamefont {Bohlin}\ \emph {et~al.}(2023)\citenamefont {Bohlin},
		\citenamefont {Turberfield}, \citenamefont {Louis},\ and\ \citenamefont
		{Šulc}}]{Bohlin2023}%
	\BibitemOpen
	\bibfield  {author} {\bibinfo {author} {\bibfnamefont {J.}~\bibnamefont
			{Bohlin}}, \bibinfo {author} {\bibfnamefont {A.~J.}\ \bibnamefont
			{Turberfield}}, \bibinfo {author} {\bibfnamefont {A.~A.}\ \bibnamefont
			{Louis}},\ and\ \bibinfo {author} {\bibfnamefont {P.}~\bibnamefont {Šulc}},\
	}\bibfield  {title} {\bibinfo {title} {Designing the self-assembly of
			arbitrary shapes using minimal complexity building blocks},\ }\href
	{https://doi.org/10.1021/acsnano.2c09677} {\bibfield  {journal} {\bibinfo
			{journal} {ACS Nano}\ }\textbf {\bibinfo {volume} {17}},\ \bibinfo {pages}
		{5387} (\bibinfo {year} {2023})}\BibitemShut {NoStop}%
	\bibitem [{\citenamefont {Pinto}\ \emph {et~al.}(2023)\citenamefont {Pinto},
		\citenamefont {Šulc}, \citenamefont {Sciortino},\ and\ \citenamefont
		{Russo}}]{Pinto2023}%
	\BibitemOpen
	\bibfield  {author} {\bibinfo {author} {\bibfnamefont {D.~E.~P.}\
			\bibnamefont {Pinto}}, \bibinfo {author} {\bibfnamefont {P.}~\bibnamefont
			{Šulc}}, \bibinfo {author} {\bibfnamefont {F.}~\bibnamefont {Sciortino}},\
		and\ \bibinfo {author} {\bibfnamefont {J.}~\bibnamefont {Russo}},\ }\bibfield
	{title} {\bibinfo {title} {Design strategies for the self-assembly of
			polyhedral shells},\ }\href {https://doi.org/10.1073/pnas.2219458120}
	{\bibfield  {journal} {\bibinfo  {journal} {Proceedings of the National
				Academy of Sciences}\ }\textbf {\bibinfo {volume} {120}},\ \bibinfo {pages}
		{e2219458120} (\bibinfo {year} {2023})}\BibitemShut {NoStop}%
	\bibitem [{\citenamefont {Romano}\ \emph {et~al.}(2020)\citenamefont {Romano},
		\citenamefont {Russo}, \citenamefont {Kroc},\ and\ \citenamefont
		{{\v{S}}ulc}}]{Romano2020}%
	\BibitemOpen
	\bibfield  {author} {\bibinfo {author} {\bibfnamefont {F.}~\bibnamefont
			{Romano}}, \bibinfo {author} {\bibfnamefont {J.}~\bibnamefont {Russo}},
		\bibinfo {author} {\bibfnamefont {L.}~\bibnamefont {Kroc}},\ and\ \bibinfo
		{author} {\bibfnamefont {P.}~\bibnamefont {{\v{S}}ulc}},\ }\bibfield  {title}
	{\bibinfo {title} {Designing patchy interactions to self-assemble arbitrary
			structures},\ }\href@noop {} {\bibfield  {journal} {\bibinfo  {journal}
			{Physical Review Letters}\ }\textbf {\bibinfo {volume} {125}},\ \bibinfo
		{pages} {118003} (\bibinfo {year} {2020})}\BibitemShut {NoStop}%
	\bibitem [{\citenamefont {E{\'{e}}n}\ and\ \citenamefont
		{Biere}(2005)}]{Een2005}%
	\BibitemOpen
	\bibfield  {author} {\bibinfo {author} {\bibfnamefont {N.}~\bibnamefont
			{E{\'{e}}n}}\ and\ \bibinfo {author} {\bibfnamefont {A.}~\bibnamefont
			{Biere}},\ }\bibfield  {title} {\bibinfo {title} {{Effective preprocessing in
				SAT through variable and clause elimination}},\ }\href
	{https://doi.org/10.1007/11499107_5/COVER/} {\bibfield  {journal} {\bibinfo
			{journal} {Lecture Notes in Computer Science}\ }\textbf {\bibinfo {volume}
			{3569}},\ \bibinfo {pages} {61} (\bibinfo {year} {2005})}\BibitemShut
	{NoStop}%
	\bibitem [{\citenamefont {Bol}(1982)}]{bol1982monte}%
	\BibitemOpen
	\bibfield  {author} {\bibinfo {author} {\bibfnamefont {W.}~\bibnamefont
			{Bol}},\ }\bibfield  {title} {\bibinfo {title} {Monte carlo simulations of
			fluid systems of waterlike molecules},\ }\href@noop {} {\bibfield  {journal}
		{\bibinfo  {journal} {Molecular Physics}\ }\textbf {\bibinfo {volume} {45}},\
		\bibinfo {pages} {605} (\bibinfo {year} {1982})}\BibitemShut {NoStop}%
	\bibitem [{\citenamefont {Kern}\ and\ \citenamefont
		{Frenkel}(2003)}]{Kern2003}%
	\BibitemOpen
	\bibfield  {author} {\bibinfo {author} {\bibfnamefont {N.}~\bibnamefont
			{Kern}}\ and\ \bibinfo {author} {\bibfnamefont {D.}~\bibnamefont {Frenkel}},\
	}\bibfield  {title} {\bibinfo {title} {{Fluid–fluid coexistence in
				colloidal systems with short-ranged strongly directional attraction}},\
	}\href {https://doi.org/10.1063/1.1569473} {\bibfield  {journal} {\bibinfo
			{journal} {The Journal of Chemical Physics}\ }\textbf {\bibinfo {volume}
			{118}},\ \bibinfo {pages} {9882} (\bibinfo {year} {2003})}\BibitemShut
	{NoStop}%
	\bibitem [{\citenamefont {Rovigatti}\ \emph {et~al.}(2018)\citenamefont
		{Rovigatti}, \citenamefont {Russo},\ and\ \citenamefont
		{Romano}}]{Rovigatti2018}%
	\BibitemOpen
	\bibfield  {author} {\bibinfo {author} {\bibfnamefont {L.}~\bibnamefont
			{Rovigatti}}, \bibinfo {author} {\bibfnamefont {J.}~\bibnamefont {Russo}},\
		and\ \bibinfo {author} {\bibfnamefont {F.}~\bibnamefont {Romano}},\
	}\bibfield  {title} {\bibinfo {title} {How to simulate patchy particles},\
	}\href {https://doi.org/10.1140/epje/i2018-11667-x} {\bibfield  {journal}
		{\bibinfo  {journal} {The European Physical Journal E}\ }\textbf {\bibinfo
			{volume} {41}},\ \bibinfo {pages} {59} (\bibinfo {year} {2018})}\BibitemShut
	{NoStop}%
	\bibitem [{\citenamefont {Noya}\ and\ \citenamefont {Doye}(2024)}]{Noya2024}%
	\BibitemOpen
	\bibfield  {author} {\bibinfo {author} {\bibfnamefont {E.~G.}\ \bibnamefont
			{Noya}}\ and\ \bibinfo {author} {\bibfnamefont {J.~P.~K.}\ \bibnamefont
			{Doye}},\ }\href {https://arxiv.org/abs/2407.17212} {\bibinfo {title} {A
			one-component patchy-particle icosahedral quasicrystal}} (\bibinfo {year}
	{2024})\BibitemShut {NoStop}%
\end{thebibliography}
\end{document}